\documentclass[journal]{IEEEtran}
\usepackage{amssymb}
\usepackage{times}
\usepackage{soul}
\usepackage{xtab}
\usepackage{url}
\usepackage{tikz}
\usetikzlibrary{mindmap,trees}
\usepackage[hidelinks]{hyperref}
\usepackage[utf8]{inputenc}
\usepackage[small]{caption}
\usetikzlibrary{positioning, shapes.multipart}
\usepackage{cite}
\usepackage{graphicx}
\usepackage{amsmath,amssymb}
\usepackage{algorithm}
\usepackage{algorithmic}
\usepackage{url}
\usepackage{hyperref}

\usepackage{graphicx}
\usepackage{tikz}
\usetikzlibrary{positioning, shapes.multipart}
\usepackage{amsmath}
\usepackage{amsthm}
\usepackage{booktabs}
\usepackage{algorithm}
\usepackage{algorithmic}
\usepackage[switch]{lineno}
\usepackage{xcolor}
\usepackage{array}
\usepackage{amsmath} 
\usepackage{amssymb} 

\usepackage[edges]{forest}
\usetikzlibrary{arrows.meta, positioning}

\newtheorem{definition}{\bf{Definition}}[section]

\newcommand{\DEL}[1]{\iffalse #1 \fi}
\newcommand{\bl}{\color{blue}}
\newcommand{\rd}{\color{black}}

\usepackage{tikz}
\usetikzlibrary{trees}

\newcommand{\qiu}[1]{\textcolor{red}{(\textbf{Qiu:~#1})}}

\title{A Decade of Metric Differential Privacy: Advancements and Applications}

\author{Xinpeng Xie, Chenyang Yu, Yan Huang, Yang Cao, and Chenxi Qiu%
\thanks{Chenxi Qiu is the corresponding author (email: chenxi.qiu@unt.edu).}
\thanks{Xinpeng Xie, Chenyang Yu, Yan Huang, and Chenxi Qiu are with the Department of Computer Science and Engineering, University of North Texas, USA (emails: \{xinpengxie, chenyangyu\}@my.unt.edu, yan.huang@unt.edu, chenxi.qiu@unt.edu).}%
\thanks{Yang Cao is with the Institute of Science, Tokyo Institute of Technology, Japan (email: cao@c.titech.ac.jp).}
}

\begin{document}

\maketitle

\begin{abstract}

\emph{Metric Differential Privacy (mDP)} extends classical differential privacy by replacing Hamming adjacency with application-aware distance metrics, enabling utility-aligned protection for structured and continuous data such as locations, trajectories, images, and text embeddings. 

This overview article synthesizes a decade of progress (2013–2024), clarifying mDP’s foundations and its connections to central and local DP, and surveying three principal mechanism families: Laplace-based (e.g., planar/geo-indistinguishability and adaptations), Exponential-Mechanism variants with metric-aware utilities (including elastic and graph metrics), and optimization-based approaches that directly minimize expected utility loss via linear programming, decomposition, and post-processing (e.g., Bayesian remapping). We organize applications across geo-privacy and spatial crowdsourcing, text and embeddings, image and voice protection, graphs and network telemetry, and federated/edge settings, and catalog common datasets and evaluation practices. We also surface open challenges, robust composition and adversarial modeling, distribution shift, context-adaptive privacy, high-dimensional scalability, and principled geometry-aware trade-off bounds, and distill practical guidance for selecting metrics, mechanisms, and metrics of utility. The goal is a unified reference and roadmap for deploying scalable, utility-preserving metric privacy in real-world systems. 
\end{abstract}

\vspace{-0.0in}
\section{Introduction}

In the era of data-driven innovation, protecting individual privacy while enabling meaningful data analysis has become a central challenge for both academia and industry. \emph{Differential Privacy (DP)} \cite{dwork2014algorithmic} has emerged as a principled standard, offering rigorous guarantees by introducing carefully calibrated randomness into query outputs. Yet, the classical DP formulation assumes a uniform notion of sensitivity across all records, defining neighboring datasets as those differing in a single entry, typically measured by Hamming distance. While effective for tabular data, this binary notion is overly restrictive and often inadequate in domains where data inhabit structured or continuous spaces.

Many real-world applications require more nuanced privacy notions that capture relationships between data points with greater fidelity. For instance, in geo-location data, revealing whether a user is within 1 kilometer versus 100 kilometers of a point conveys vastly different levels of sensitive information \cite{imola2022balancing}. Similarly, in text and image analysis, semantic or perceptual similarity often matters more than strict binary differences. Classical DP, which treats all records as equally distinct, fails to capture such subtleties. To address this gap, \emph{Metric Differential Privacy (mDP)} \cite{Chatzikokolakis-PETS2013} generalizes DP by replacing the Hamming distance with application-specific metrics. By incorporating domain-aware distance measures, mDP enhances both the flexibility and expressiveness of DP, enabling its use across domains such as geo-location obfuscation \cite{Andres-CCS2013}, word embeddings \cite{feyisetan2021private}, speech processing \cite{han2020voice}, and image privacy \cite{fan2019practical,chen2021perceptual}.

\begin{figure}
    \centering
    \includegraphics[width=0.94\linewidth]{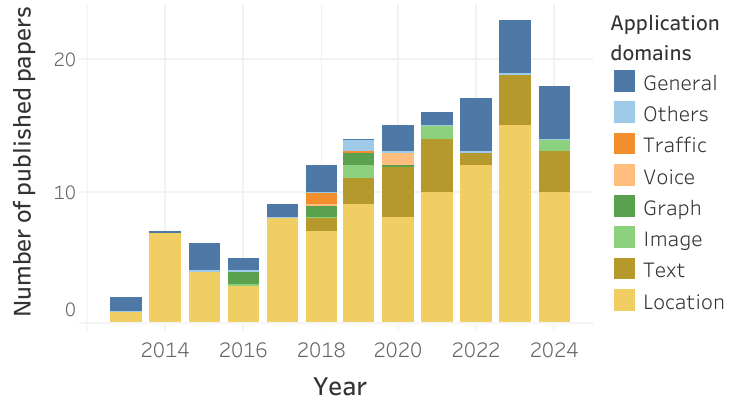}
    \caption{Growing adoption of mDP across multiple application domains.}
    \vspace{-0.00in}
    \label{fig:mDPhistory}
    \vspace{-0.00in}
\end{figure}

Since its introduction in 2013, mDP has attracted significant research attention. Figure \ref{fig:mDPhistory} illustrates the historical growth of publications from 2013 to 2024, with stacked bars showing contributions across domains including location, text, image, and voice. The trend underscores the rising influence and broadening adoption of mDP, as early dominance in location privacy gradually gave way to more diverse applications such as text and image analysis, voice protection, and graph-based privacy. This evolution highlights mDP’s expanding impact and versatility in addressing privacy needs across modern data ecosystems.

Despite substantial progress in the theoretical foundations, algorithmic advances, and practical applications of mDP, no comprehensive overview article dedicated specifically to mDP currently exists. Existing surveys primarily address classical differential privacy and local differential privacy (LDP), touching on mDP only briefly. For example, \cite{zhao2022survey} reviews DP and LDP techniques for protecting unstructured data such as images, audio, and text, with a focus on utility challenges and trade-offs, while mentioning mDP only in passing. Similarly, \cite{zhao2024scenario} examines scenario-specific adaptations of DP for local and statistical data privacy, again treating mDP as a minor topic. These works overlook the distinctive theoretical formulations, mechanism designs, and application-specific innovations that define mDP. This gap underscores the need for a dedicated, systematic review of developments and challenges unique to mDP.

To address this gap, we present the first in-depth overview article of mDP, covering the period from 2013 to 2024. Our contributions are fourfold:
\begin{itemize}
    \item We clarify how mDP extends DP and LDP by incorporating flexible distance metrics, enabling privacy definitions that more faithfully capture semantic and geometric relationships between data records.
    \item We categorize and compare major mDP mechanisms, including \emph{Laplace}-based, \emph{Exponential mechanism}-based, and \emph{optimization}, based approaches, analyzing their theoretical guarantees, computational complexity, and suitability for different use cases.
    \item We provide a comprehensive review of mDP applications across diverse domains, such as geo-location privacy, text embeddings, image protection, and speech data, highlighting the privacy–utility trade-offs in each setting.
    \item We summarize commonly used datasets in mDP research and outline promising future directions, including context-aware formulations, scalable optimization techniques, and new metrics for evaluating privacy–utility trade-offs.
\end{itemize}

The objective of this overview article is to serve as a foundational reference for researchers, practitioners, and system designers seeking to navigate the evolving landscape of metric-based privacy protection and contribute to the next generation of privacy-preserving technologies.

The remainder of the paper is organized as follows. We begin by introducing how mDP extends from traditional global and local DP in \textbf{Section \ref{sec:DP2mDP}}, underscoring mDP's unique challenges. We then introduce the three primary mDP mechanisms, including the \emph{Laplace mechanism}, \emph{Exponential mechanism}, and \emph{optimization-based approaches} in \textbf{Section \ref{sec:mechanism}}. Next, we introduce the application domains of mDP in \textbf{Section \ref{sec:applications}} and identify current hurdles and potential future research directions in \textbf{Section \ref{sec:future}}. Finally, we make a conclusion in \textbf{Section \ref{sec:conclusions}}.

\vspace{-0.0in}
\section{From DP to mDP}



\label{sec:DP2mDP}
In its original form  \cite{Dwork-TC2006}, DP employs data perturbation to ensure a consistent level of indistinguishability for queries across ``neighboring databases'', defined as databases that differ by only a single entry (i.e., with a \emph{Hamming distance} of one). mDP, which is also called \emph{Lipschitz privacy} \cite{koufogiannis2015optimality}, $d_\mathcal{X}$-privacy \cite{feyisetan2020privacy}, and \emph{smooth DP} \cite{dharangutte2023integer}, expand this concept to consider distances between records within a general metric space, rather than restricting it to the Hamming distance.

We model a randomized mechanism as a \emph{probabilistic mapping} $\mathcal{M}: \mathcal{X} \rightarrow \mathcal{Y}$ from a \emph{secret (input) domain} $\mathcal{X}$ to a \emph{perturbed (output) domain} $\mathcal{Y}$. The input space $\mathcal{X}$ is equipped with a distance function $d_{\mathcal{X}}: \mathcal{X}^2 \rightarrow \mathbb{R}$; for any $x,x'\in\mathcal{X}$ we write $d_{\mathcal{X}}(x,x')$ for their distance.

\begin{definition}
\label{def:metricDP}
\textbf{$(\epsilon, d_{\mathcal{X}})$-mDP.}
A randomized mechanism $\mathcal{M}: \mathcal{X} \rightarrow \mathcal{Y}$ satisfies $(\epsilon, d_{\mathcal{X}})$-mDP if, for every subset $\mathcal{Y}' \subseteq \mathcal{Y}$ and all $x,x'\in\mathcal{X}$,
\vspace{-0.00in}
\begin{equation}
\label{eq:DP}
\frac{\Pr[\mathcal{M}(x) \in \mathcal{Y}']}{\Pr[\mathcal{M}(x') \in \mathcal{Y}']} \le e^{\epsilon\, d_{\mathcal{X}}(x,x')}, \quad \forall\, \mathcal{Y}' \subseteq \mathcal{Y},
\vspace{-0.00in}
\end{equation}
where $\epsilon>0$ is the \textbf{privacy budget}.
\end{definition}

Intuitively, mDP ensures that \emph{nearby inputs (under $d_{\mathcal{X}}$) induce output distributions that differ by at most an $e^{\epsilon d_{\mathcal{X}}(x,x')}$ multiplicative factor}. Thus, small input changes lead to only bounded changes in the law of $\mathcal{M}(x)$.

\DEL{
\vspace{0.05in}
\noindent \textbf{Threat model.}
We consider a Bayesian adversary who observes perturbed outputs and updates a prior belief over the secret $x\in\mathcal{X}$. Following Kerckhoffs’ principle, the adversary knows the mechanism family and parameters: the mapping $\mathcal{M}$, the metric $d_{\mathcal{X}}$, and the privacy budget $\epsilon$. The adversary may also hold contextual side information $S$ (e.g., road networks, feasible speeds, POIs, time-of-day patterns, or population mobility statistics) and an arbitrary prior $\pi$ on $\mathcal{X}$, which can be population-level (learned from public traces) or personalized (learned from the same user’s history). 

\emph{Observations.}
In the single-shot setting, the adversary observes $y=\mathcal{M}(x)$. More generally, the observation can be any (possibly coarsened) function of the output (e.g., rounded coordinates, a top-$k$ list, or server logs). In the sequential setting, the adversary observes a (linkable) sequence $\mathbf{y}_{1:T}=(y_1,\dots,y_T)$ with timestamps, corresponding to secrets $\mathbf{x}_{1:T}$. We consider both honest-but-curious service providers and passive eavesdroppers; active probing of the service is allowed but does not weaken the mDP guarantees.

\emph{Goals.}
Typical objectives include point estimation (recover $x$ via MAP), region/property inference (e.g., membership in a sensitive region), and trajectory reconstruction. Given a known mechanism $\mathcal{M}$ and prior $\pi$, the posterior is (by Bayes’ rule)~\cite{Yu-NDSS2017}
\vspace{-0.05in}
\begin{equation}
\label{eq:bayes}
\Pr[X=x \mid Y=y]
= \frac{\pi(x)\,\Pr[\mathcal{M}(x)=y]}{\sum_{x'\in\mathcal{X}} \pi(x')\,\Pr[\mathcal{M}(x')=y]}.
\end{equation}
\vspace{-0.00in}
The \emph{maximum a posteriori} estimate is
\vspace{-0.00in}
\begin{equation}
\label{eq:bayes-map}
\hat{x} \;=\; \arg\max_{x\in\mathcal{X}} \Pr[X=x \mid Y=y].
\end{equation}

\emph{Privacy guarantee under mDP.}
Under mDP, for any $x,x'\in\mathcal{X}$ and any $y\in\mathcal{Y}$,
$\frac{\Pr[\mathcal{M}(x)=y]}{\Pr[\mathcal{M}(x')=y]}\le e^{\epsilon d_{\mathcal{X}}(x,x')}$.
Plugging this into Eq.~\eqref{eq:bayes} yields the \emph{posterior-odds} bound
\begin{eqnarray}
\frac{\Pr[X=x \mid Y=y]}{\Pr[X=x' \mid Y=y]}
&=& \frac{\pi(x)}{\pi(x')} \cdot
\frac{\Pr[Y=y \mid X=x]}{\Pr[Y=y \mid X=x']} \\
&\leq&
\frac{\pi(x)}{\pi(x')} \cdot e^{\epsilon d_{\mathcal{X}}(x,x')}.
\end{eqnarray}
Hence, mDP limits how much observing $y$ can tilt the adversary’s posterior odds between any two secrets, in proportion to their distance, \emph{for any prior $\pi$ and any side information $S$ captured in the observation model}.

\emph{Sequential composition.}
If releases at times $t=1,\dots,T$ use mechanisms each satisfying $(\epsilon_t,d_{\mathcal{X}})$-mDP (conditioned on $X_t$), then the joint release satisfies $\big(\sum_t \epsilon_t, d_{\mathcal{X}}\big)$-mDP; equivalently, posterior-odds scale by $\exp((\sum_t \epsilon_t)\, d_{\mathcal{X}}(x,x'))$. For trajectory secrets $\mathbf{x}_{1:T}$, an additive path metric $D(\mathbf{x}_{1:T},\mathbf{x}'_{1:T})=\sum_t d_{\mathcal{X}}(x_t,x'_t)$ yields the analogous bound with $D$.

\emph{Scope.}
Unless explicitly modeled as part of the observation, we exclude external side channels (e.g., battery, radio beacons, payment events). Such channels can be incorporated into $S$ or the observation map without invalidating the mDP inequality, though they may induce a sharper prior $\pi$ and thus strengthen attacks within the same guarantee.

}

\vspace{0.05in}
\noindent \textbf{mDP vs. local DP.} mDP defined in \textbf{Definition \ref{def:metricDP}} is closely related to the definition of \emph{local DP} \cite{KasiviswanathanJoC}, where a random mechanism $\mathcal{M}$ is applied independently to each element of a database. The distinction between mDP and local DP lies in the inclusion of the distance metric $d_{\mathcal{X}}(x, x')$ in the mDP formulation, which is absent in local DP. This inclusion requires the perturbation mechanism to ensure indistinguishability for elements $x, x' \in \mathcal{X}$ based on their distance $d_{\mathcal{X}}(x, x')$, demanding a more precise control of data perturbation. In addition, mDP retains its properties under both \emph{post-processing} and the \emph{sequential composition} of mechanisms like traditional DP \cite{imola2022balancing}. 

\begin{table*}[t]
\scriptsize    
\centering
\begin{tabular}{p{4.8cm}|p{2.2cm}|p{9.5cm}}
\toprule
\textbf{Paper} & \textbf{Distance Type} & \textbf{Application: Description} \\ \hline
\hline
PETS 2013 \cite{Chatzikokolakis-PETS2013}, PETS 2014 \cite{chatzikokolakis2014predictive}, CCS 2014 \cite{fawaz2014location}, UAI \cite{imola2022balancing}, ICDM 2019 \cite{feyisetan2019leveraging} & Euclidean Distance & Location-based services (LBS) and natural language processing (NLP): $d_{\mathcal{X}}(x, x') = \sqrt{\sum_{i}(x_i - x'_i)^{2}}$ quantifies the straight-line displacement between two data points $x, x'$, where $x_i$ and $x'_i$ represent the $i$-th coordinate of $x$ and $x'$, respectively \\ \hline

CDC 2014 \cite{wang2014entropy} & Manhattan Distance & Control Systems: $d_{\mathcal{X}}(x, x') = \sum_{i}|x_i - x'_i|$ measures the total absolute difference across all coordinate dimensions. \\ \hline

arXiv 2018 \cite{fernandes2018author}, POST 2019 \cite{Fernandes-PST2019} & Word Mover's Distance (Kantorovich Distance) & NLP: $d_{\mathcal{X}}(x, x')$ measures the minimum transport cost needed to move words from one text document $x$ to another $x'$. The cost is computed based on the semantic similarity of the words, measured using distances between their vector representations in a shared embedding space.\\ \hline

FOCS 2018 \cite{borgs2018revealing} & Node-Distance & Graph processing: $d_{\mathcal{X}}(x, x')$ measures the minimum number of node modifications such as additions, deletions, or substitutions needed to transform one graph structure $x$ into another $x'$. \\ \hline

VLDB 2015 \cite{haney2014design}, TMC 2020 \cite{qiu2020location}, ESORICS 2020 \cite{cao2020pglp}, TITS 2022 \cite{ma2022personalized} & Shortest Path Distance & LBS: $d_{\mathcal{X}}(x, x')$ measures the minimum travel distance between two points $x, x'$ along a network, considering the road or graph structure connecting them. \\ \hline


ESORICS 2019 \cite{kawamoto2019local} & Wasserstein Distance & LBS: $d_{\mathcal{X}}(x, x')$ measures the minimal cost of transforming one probability distribution $x$ into another $x'$, based on optimal transport theory. \\ \hline

arXiv 2020 \cite{xu2020differentially}, TKDE 2023 \cite{zhao2022geo} & Regularized Mahalanobis Norm & LBS and NLP: $d_{\mathcal{X}}(x, x')$ measures distance between $x$ and $x'$ while considering correlations between their features, regularized to balance sensitivity and robustness in privacy mechanisms. \\ \hline

ICME 2020 \cite{han2020voice}, CCS 2021 \cite{weggenmann2021differential}, ESORICS 2021 \cite{fernandes2021locality} & Angular Distance & Voice processing and LBS: $d_{\mathcal{X}}(x, x')$ measures the difference in orientation between vectors $x, x'$, often using cosine similarity to quantify differences in feature representations. \\ \hline

SEBD 2023 \cite{boninsegna2023locality} & Fréchet Distance & LBS: $d_{\mathcal{X}}(x, x')$ measures the similarity between two curves $x, x'$ by considering the minimal effort required to continuously transform one into the other while maintaining order. \\ \hline

arXiv 2024 \cite{brauer2024time} & Temporal Distance & LBS: $d_{\mathcal{X}}(x, x')$ measures the difference between time points or temporal sequences $x$ and $x'$,  capturing the deviation in timestamps along trajectories. \\ 
\toprule
\end{tabular}
\vspace{-0.00in}
\caption{Metric distance definitions and descriptions in mDP works. 
}
\label{tab:mDP-distances}
\vspace{-0.00in}
\end{table*}

Compared to local DP, mDP offers several \textbf{advantages}:

\emph{(1) Enhanced utility for high-dimensional data}: Local DP typically incurs significant utility loss on high-dimensional or continuous data, since preserving privacy requires adding noise that scales exponentially with dimension \cite{yang2024local}. In contrast, mDP leverages distance metrics to inject noise adaptively, based on the similarity between data points. By limiting indistinguishability primarily among nearby points, mDP preserves structural patterns in the original data, maintaining higher utility for downstream tasks. This makes mDP particularly effective for applications such as image data, audio signals, and text embeddings \cite{fawaz2014location}.

\emph{(2) Context-aware privacy mechanisms}: mDP enables the design of mechanisms aligned with domain-specific distance metrics. For geographic data, Euclidean or haversine distance naturally preserves spatial relationships \cite{Andres-CCS2013}, while in text data, Word Mover’s Distance captures semantic similarity between documents \cite{fernandes2018author}. Such context-aware criteria allow the mechanisms to reflect intrinsic data properties, offering stronger alignment between privacy and utility than the uniform randomization in local DP \cite{chen2021perceptual}. 

Table \ref{tab:mDP-distances} summarizes commonly used distance metrics in mDP-related work, ranging from classical geometric measures (e.g., Euclidean, Manhattan, and shortest path) to semantic metrics (e.g., Word Mover’s Distance and Wasserstein) and more specialized notions for structural or temporal data (e.g., node-distance for graphs, angular distance for embeddings and voice, Fréchet distance for trajectories, and temporal distance for timestamped sequences). Together, these metrics highlight the flexibility of mDP in aligning privacy guarantees with meaningful, domain-specific notions of similarity across spatial, textual, structural, and perceptual data.

\emph{(3) Fine-grained privacy–utility trade-offs}: Unlike DP’s rigid “one-record difference” notion, mDP naturally accommodates continuous data types such as geographic coordinates, audio embeddings, or image features. By tuning the noise relative to the chosen distance metric, practitioners gain more nuanced control over the privacy–utility balance. This flexibility enables application-specific optimization of trade-offs that local DP’s uniform mechanisms often cannot achieve \cite{imola2022balancing}.




\DEL{

Having defined the principles and advantages of mDP in Section \ref{sec:DP2mDP}, we now introduce how mDP protect against a range of attack models, including xxx, xxx, xxx. Table~\ref{tab:mdp-attacks-final} provides a summary of representative attack models that have been addressed in the mDP literature.

{\bl It is still unclear how the attack models are related to mDP in the following models. Further discussion is needed. }

{\rd More advanced location-based attacks build upon this foundation. For instance, the Markov Chain Attack leverages temporal correlations, assuming mobility follows a Markov model, to infer an entire trajectory by maximizing the joint posterior probability of a location sequence~\cite{cao2020pglp}. Other sophisticated threats include the Traffic Flow Aware Inference Attack, which uses public traffic data to de-obfuscate vehicle locations~\cite{qiu2022trafficadaptor}, the Profile-Estimation Based Attack, which learns a user's mobility profile to improve inference~\cite{oya2019rethinking, mendes2020impact}, and the Map-Matching Attack, which uses road network constraints to identify a vehicle's most likely path~\cite{kubicka2018comparative, mendes2020impact, cunha2019clustering}.

Beyond location data, several attack models target the identity of individuals or the content of their data. The Re-identification Attack is a general threat where an adversary attempts to link anonymized data records back to specific individuals using background knowledge. A notable variant is the Top-$N$ Re-identification Attack, where the adversary's goal is to include the user's true location within a ranked list of $N$ candidates that maximize the cumulative posterior probability~\cite{duarte2024privacy}. This is a critical threat in scenarios from trajectory analysis to facial recognition~\cite{zang2011anonymization, yan2024coder}. In the text domain, a Deanonymization Attack specifically aims to determine the authorship of a document by analyzing stylistic or semantic features that persist even after perturbation mechanisms have been applied~\cite{utpala2023locally, Mattern2022Limits}.

With the rise of machine learning, attacks targeting models and their training data have become increasingly relevant. A Reconstruction Attack aims to recover original data points from the noisy output of a privacy mechanism, which can be achieved by applying Bayesian inference to each perturbed report~\cite{odoh2024group, dwork2017exposed, arnold2023guiding}. A more advanced threat in collaborative learning is the Deep Leakage from Gradients (DLG) Attack, where an adversary can perfectly reconstruct private training data by observing shared gradients. The attack works by finding a ``dummy'' input \( (\hat{x}, \hat{y}) \) whose gradients match the true gradients~\cite{galli2023group, galli2023advancing}:

\begin{equation}
(\hat{x}, \hat{y}) = \arg\min_{(x', y')} \left\| \nabla \mathcal{L}(\theta; x', y') - \nabla \mathcal{L}(\theta; x, y) \right\|_2^2.
\end{equation}

A related threat is the Membership Inference Attack, where the goal is to determine whether a specific individual's data was part of a model's training set, posing a significant privacy risk~\cite{yan2024coder, shokri2017membership, arnold2023guiding, carvalho2023tem}. Finally, the applicability of mDP extends to other data modalities, each with its own threat models, such as the Spoofing Attack, which attempts to identify individuals from their voiceprints even after privacy-preserving transformations have been applied~\cite{han2020voice}.}}

\vspace{-0.00in}
\section{Key Data Perturbation Mechanisms in mDP}
\label{sec:mechanism}
\vspace{-0.00in}

\begin{figure*}
    \centering
    \includegraphics[width=0.98\linewidth]{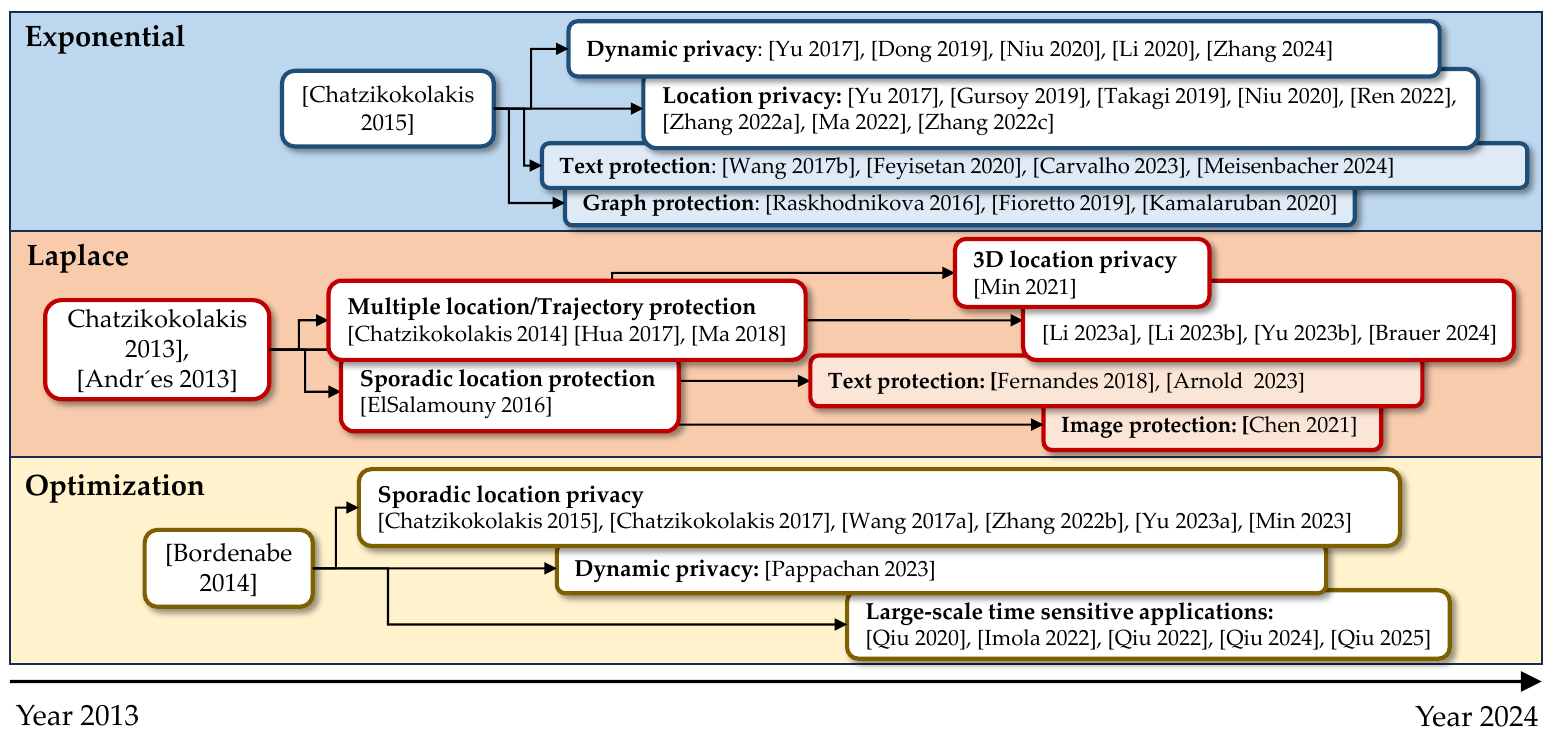}
    \vspace{-0.00in}
    \caption{The evolvement of the three categories of mDP mechanisms during years 2013 -- 2024 (due to the limited space, we label each work by [Last name of the first author, publication year].}
    \label{fig:relatedwork}
    \vspace{-0.00in}
\end{figure*}

In this section, we review the core mechanisms that have shaped the development of mDP: the Laplace mechanism (\textbf{Section \ref{subsec:laplace}}), the Exponential mechanism (\textbf{Section \ref{subsec:exponential}}), and optimization-based approaches (\textbf{Section \ref{subsec:optimization}}). A comparative analysis of these mechanisms is presented in \textbf{Section \ref{subsec:comparison}}. The historical evolution of these three categories is illustrated in Figure \ref{fig:relatedwork}.

\DEL{
\begin{table*}[ht]
\centering
\caption{Representative attack models addressed in mDP literature}
\label{tab:mdp-attacks-final}
\begin{tabular}{p{3.0cm}|p{9.0cm}|p{4.5cm}}
\toprule
\textbf{Attack Model} & \textbf{Description} & \textbf{Targeted Domain} \\
\hline
\hline
Bayesian Inference Attack \cite{Yu-NDSS2017,qiu2024enhancing,takagi2023geo} & Attacker uses Bayesian inference to refine location estimates from obfuscated data, aiming to minimize expected error. & Location Data, Spatial Crowdsourcing, LBS \\
\hline
Deanonymization Attack \cite{utpala2023locally,Mattern2022Limits} & Re-identifies a document's author via stylistic or semantic features that persist despite perturbation. & Text, Author Attribution \\
\hline
Reidentification Attack \cite{duarte2024privacy,zang2011anonymization,yan2024coder} & Links anonymized data to individuals using background knowledge. Examples: matching Top-N locations or using facial recognition on perturbed images. & Location, Trajectory, LBS, Image Data, Face Recognition \\
\hline
Reconstruction Attack \cite{odoh2024group,dwork2017exposed,arnold2023guiding} & Recovers original records from the noisy output of a privacy mechanism. For text, may involve finding the nearest neighbor to a substitute word. & General Data, Statistical Databases, Text Data, Language Models \\
\hline
Membership Inference Attack \cite{shokri2017membership,yan2024coder,arnold2023guiding,carvalho2023tem} & Determines if a specific data record (e.g., an image or text) was part of a machine learning model's training dataset. & Image Data, Text Data, Machine Learning Models \\
\hline
Deep Leakage from Gradients Attack \cite{zhu2019deep,galli2023group,galli2023advancing} & Reconstructs private training data in collaborative learning by matching the gradients of a dummy input to the observed gradients. & Federated Learning, Collaborative Machine Learning, Gradients \\
\hline
Traffic Flow Aware Inference Attack \cite{qiu2022trafficadaptor} & Leverages public vehicle traffic flow data (often with Hidden Markov Models) to infer true locations from obfuscated reports. & Vehicle Location Data, Trajectory Data, LBS, Road Networks, Traffic Flow \\
\hline
Profile-Estimation Based Attack \cite{oya2019rethinking,mendes2020impact} & Estimates a user's mobility profile from a series of obfuscated reports to improve subsequent location inference. & Location Data, Mobility Traces, LBS \\
\hline
Map-Matching Attack \cite{kubicka2018comparative,mendes2020impact,cunha2019clustering} & Uses road network constraints (often with HMMs) to identify a vehicle's most plausible path from noisy location data. & Vehicle Location Data, Trajectory Data, Road Networks, LBS \\
\hline
Spoofing Attack \cite{wu2015spoofing,han2020voice} & Identifies or verifies a speaker from protected speech data by analyzing unique physiological and behavioral voice characteristics. & Speech Data, Voiceprint, Speaker Recognition, Biometric Identification \\
\bottomrule
\end{tabular}
\end{table*}
}

\subsection{Laplace Mechanism}
\label{subsec:laplace}
The Laplace mechanism is one of the most widely used data perturbation techniques, originally developed to achieve Differential Privacy (DP) \cite{Dwork-TC2006}. It adds noise $w$ to the output of a query function $f(x)$, i.e.,
$\mathcal{M}(f(x)) = f(x)+w$, where $w$ is sampled from a Laplace distribution with a probability density function
\vspace{-0.05in}
\begin{equation}
\Pr\left[\mathcal{M}(f(x)) = f(x)+w\right] 
= \frac{\epsilon}{2\Delta f} e^{-\frac{\epsilon |w|}{\Delta f}}.
\vspace{-0.00in}
\end{equation}
Here, $\Delta f = \sup_{x, x' \in \mathcal{X}, d_{\mathrm{h}}(x, x')=1} |f(x) - f(x')|$ represents the global sensitivity of $f$, which measures the maximum change in the function's output between neighboring datasets ($d_{\mathrm{h}}(x, x')$ denotes the Hamming distance between $x$ and $x'$).

A pivotal extension of this concept is metric-based differential privacy, or $d_{\mathcal{X}}$-privacy, which generalizes the notion of adjacency from Hamming distance to an arbitrary metric $d_{\mathcal{X}}$ \cite{Chatzikokolakis-PETS2013}. This framework allows the mechanism to be tailored to specific data types and privacy requirements by defining a more nuanced ``distance'' between data records. The Laplace mechanism can naturally adapt to various metrics by appropriately scaling the noise \cite{McSherry-FOCS2007}. To extend Laplace for mDP, \cite{Chatzikokolakis-PETS2013} first proposed scaling the noise based on the metric $d_{\mathcal{X}}$ instead of the global sensitivity $\Delta f$, where $d_{\mathcal{X}}(x, y)$ quantifies the distinguishability between any two records $x, y\in \mathcal{X}$. The resulting probability density function is expressed as
\vspace{-0.00in}
\begin{equation}
\Pr\left[\mathcal{M}(x) = y\right] = \lambda e^{-d_{\mathcal{X}}(x, y)}.
\vspace{-0.00in}
\end{equation}
Following this work, \cite{Andres-CCS2013} introduced polar Laplace noise to achieve \emph{geo-indistinguishability (Geo-Ind)}. This approach employs a two-dimensional Laplace distribution centered on the true location $x$, with a probability function given by
$$
\Pr\left[\mathcal{M}(x) = y\right] = \frac{\epsilon^2}{2\pi} e^{-\epsilon \|x - y\|_2}.
$$
This foundational work paved the way for a rich body of research applying and extending the Laplace mechanism in numerous contexts \cite{Andres-CCS2013, hua2017geo, ma2018agent, Chen-IoTJ2022, mendes2018effect, to2018privacy, yang2021blockchain, huang2019incentivizing, li2023privacy, li2023accurate, duarte2024privacy, min2023personalized, yu2023privacy, al2018adaptive}.

A major challenge in applying these mechanisms lies in protecting continuous location traces, where the privacy budget can be rapidly exhausted. To mitigate this, early work introduced a \emph{predictive mechanism} that exploits temporal correlations in mobility data: rather than injecting fresh noise at every time step, it reuses previously generated noise when the user’s location is predictable, thereby conserving the privacy budget \cite{chatzikokolakis2014predictive}. Subsequent research extended this idea by examining how factors such as update frequency affect privacy and utility \cite{mendes2018effect, al2018adaptive}, and by developing more advanced techniques. Examples include clustering trajectories to obfuscate only stay-point centers \cite{cunha2019clustering} and perturbing key parameters of trajectory segments instead of all individual points \cite{li2023privacy}. The urgency of protecting location traces became especially evident during the COVID-19 pandemic, when Laplace-based mechanisms were employed for privacy-preserving contact tracing and vulnerability mapping \cite{li2023accurate, Chen-IoTJ2022}. More recently, these approaches have been extended to protect entire “longitudinal” trajectories in applications such as location-based advertising \cite{yu2023privacy}.

In parallel with these practical extensions, a significant body of work has focused on a deeper theoretical understanding of the Laplace mechanism itself, particularly its optimality and utility. Early work established its optimality from an entropy-minimization perspective for discrete-time systems \cite{wang2014entropy}. This was followed by rigorous proofs showing that the Laplace mechanism minimizes mean-squared error for identity queries under a slightly stronger notion called Lipschitz privacy \cite{koufogiannis2015optimality}. This optimality was ultimately proven to hold even for the challenging case of continuous queries over continuous domains, solidifying its foundational status \cite{fernandes2021laplace}. Despite its optimality, its utility is not always sufficient. Critical analysis has shown that the privacy-utility trade-off can be unfavorable, requiring very high noise for a meaningful privacy guarantee \cite{oyaGeoIndLooking}. This has spurred research into utility improvement techniques, such as post-processing ``remapping'' which uses prior knowledge to move obfuscated points from unlikely regions (e.g., a lake) to more plausible ones \cite{chatzikokolakis2017efficient}, a concept later refined to ensure the remapping process itself is private \cite{duarte2024privacy}. Further theoretical work has focused on making the privacy guarantees more tangible by interpreting the parameter $\epsilon$ in terms of concrete adversarial advantage \cite{pankova2022interpreting}. Another fascinating theoretical development showed that it is possible to ``gradually release'' data, starting with a high level of Laplace noise and later revealing a more accurate version without any utility loss compared to releasing the accurate version from the start \cite{Koufogiannis-JPC2017}.

Recognizing that a ``one-size-fits-all'' approach is often suboptimal, researchers have developed numerous ways to customize and adapt the Laplace mechanism. One major direction is making the privacy level itself dependent on the data. ``Location-dependent privacy'' formalizes this by making the privacy parameter $\epsilon$ a function of the location itself, providing stronger protection in more sensitive areas, a concept linked computationally to the Eikonal equation \cite{koufogiannis2016location}. Practical implementations of this idea include ``tiled mechanisms'' that assign different privacy levels to different geographical tiles \cite{AhujaEDBT2019}, and ``semantic adaptive'' mechanisms that adjust noise based on a location's meaning \cite{min2024semantic}. The mechanism has also been adapted for specific physical domains, most notably for 3D indoor environments, requiring a 3D Laplace distribution and specialized sampling techniques to respect building geometry \cite{min20213d, min2023personalized}. Some have proposed a ``truncated'' geo-indistinguishability for indoor spaces to ensure obfuscated points remain within the building \cite{yang2021blockchain, fathalizadeh2023indoor}. For further customization, frameworks like ``Blowfish'' and \emph{Policy Graph based Location Privacy (PGLP)} use a ``policy graph'' to define indistinguishability, moving beyond simple Euclidean distance to capture application-specific privacy needs, such as those in ranked geo-queries \cite{haney2014design, cao2020pglp, eltarjaman2017location, kamalaruban2020not}.

The versatility of the Laplace mechanism's core principle is most evident in its wide-ranging application to non-spatial data. In natural language processing, this principle has been used for author obfuscation by adding noise to stylometric feature vectors \cite{fernandes2018author} and for privatizing text by replacing words based on semantic or syntactic similarity, guided by metrics over word embeddings or grammatical structures \cite{Fernandes-PST2019, feyisetan2019leveraging, arnold2023guiding}. A more direct application involves the private release of the high-dimensional text embedding vectors themselves \cite{feyisetan2021private}. The principle has also been generalized to protect entire probability distributions using the Wasserstein distance as the metric \cite{kawamoto2019local} and to add noise to graph data to protect critical infrastructure networks \cite{fioretto2019privacy} or to privately estimate network model parameters \cite{borgs2018revealing}. Other novel applications include protecting time series data from IoT sensors \cite{fan2018time} and even obfuscating facial images in a GAN's latent space to achieve ``perceptual indistinguishability'' \cite{chen2021perceptual}. In complex systems like ridesharing, mobile crowdsourcing, and federated learning, the Laplace mechanism or its underlying principles are often key components for protecting locations during task assignment \cite{tong2017jointly, to2018privacy, huang2016eppd} or enabling privacy-preserving incentive mechanisms \cite{huang2019incentivizing, shi2021clap}. While federated learning applications often use Gaussian noise in practice for technical reasons, the Laplace mechanism is consistently referenced as the conceptual foundation for providing rigorous differential privacy guarantees \cite{galli2023group, galli2023advancing}.

\subsection{Exponential Mechanism}
\label{subsec:exponential}

The \emph{Exponential Mechanism (EM)}, another widely used method for ensuring DP \cite{McSherry-FOCS2007}, has since been adapted to mDP \cite{chatzikokolakis2015constructing}. Unlike Laplace, which adds noise directly to the data values, EM selects an output probabilistically from \emph{a discrete set} based on a quality score. This makes it particularly well-suited for applications that aim to choose an element from a finite set rather than perturbing continuous data. Its adaptability became particularly crucial with the advent of metric differential privacy (mDP), which extends privacy guarantees to general metric spaces. The foundational adaptation of EM for mDP was presented by \cite{chatzikokolakis2015constructing}, which introduced an elastic distinguishability metric that adjusts the notion of distance based on the local density of the data space. This allows the mechanism to provide stronger protection in sparse regions and higher utility in dense ones. Formally, given an input $x \in \mathcal{X}$, the probability of selecting an output $y \in \mathcal{Y}$ is defined as:
\begin{equation}
\label{eq:exp_mech}
\Pr\left[\mathcal{M}(x) = y\right] = a_x e^{-\frac{1}{2}\epsilon d(x, y)},
\end{equation}
where $d(x, y)$ is the underlying metric, and $a_x = \left( \sum_{y' \in \mathcal{Y}} e^{-\frac{1}{2}\epsilon d(x, y')} \right)^{-1}$ is the normalization constant. The utility function is implicitly the negative distance $-d(x,y)$, ensuring that outputs closer to the input are exponentially more likely. This metric-aware formulation of EM laid the groundwork for a wide range of sophisticated privacy-preserving techniques.

Following this work, EM has been extensively adapted for location privacy. A crucial advancement was tailoring the mechanism for the unique constraints of ``road networks'', where Euclidean distance is often a poor proxy for actual travel distance. A significant step forward was the proposal of \emph{Geo-Graph-Indistinguishability (GG-I)}, which uses the shortest-path distance on a road network graph as the metric within the EM framework \cite{takagi2023geo}. This ensured that perturbed locations were not only topologically close but also always on the road network. This idea was further evolved by \cite{ma2022new}, which introduced a \emph{Cloaking Region Obfuscation (CRO)} mechanism that uses EM to obfuscate an entire region to another, allowing for cooperative privacy among vehicles. For example, \cite{Yu-NDSS2017} developed the PIVE framework, which combines Geo-Ind with another privacy criterion, \emph{expected inference error (EIE)}, to dynamically adjust privacy guarantees based on user-defined thresholds \cite{shun2021differential, zhang2024dpive}. \cite{gursoy2019secure} extended EM to \emph{condensed local DP} mechanisms for datasets involving a small population. In this adaptation, similar outputs are systematically prioritized over distant ones by applying condensed probability during the perturbation process. To address the problem of distributional shifts caused by privacy mechanisms, \cite{ren2022distpreserv} designed a specialized EM utility function to preserve the statistical consistency of the original data.

Meanwhile, \emph{dynamic EM} adaptations have also been designed to address evolving privacy challenges. For example, \cite{dong2019preserving} integrated EM into dynamic spectrum sharing frameworks to protect the privacy of primary users under Geo-Ind constraints during auction-based allocations. \cite{niu2020eclipse} developed an EM-based method called Eclipse, which dynamically modulates privacy protections to mitigate risks posed by long-term observation attacks. Another line of work addressed the issue of the mechanism itself revealing the user's privacy-seeking behavior. The \emph{Perturbation-Hidden} privacy model was proposed to ensure that all outputs of the EM are geographically plausible, preventing an adversary from inferring that a privacy mechanism is in use simply by observing an impossible location, like a vehicle in a lake \cite{li2020perturbation}. Furthermore, to handle location data with clear directional patterns, such as traffic on a highway, an \emph{Ellipse-based EM} was introduced to provide more utility by aligning the privacy protection area with the data's natural distribution \cite{zhao2022geo}.

EM's adaptability has also proven important for textual and graph data privacy. For textual data, \cite{wang2017local} proposed the Subset Exponential Mechanism (SEM), which tailors EM for ordinal data by incorporating a structured probability assignment. \cite{feyisetan2020privacy} leveraged EM to perturb text embeddings, ensuring privacy protection without compromising semantic coherence. To improve upon this, \cite{yue2021differential} proposed a mechanism to distinguish between sensitive and non-sensitive words, providing tailored protection. The broader challenges in this domain, such as calibrating noise to local embedding density, remain an active area of research \cite{feyisetan2020research}.

For graph data, EM has been effectively applied to protect the privacy of critical infrastructure networks. For example, \cite{raskhodnikova2016lipschitz} generalized EM to node-private graph statistics by employing Lipschitz extensions, enabling the handling of sensitivity variations in diverse graph queries, including subgraph counts and degree distributions. \cite{fioretto2019privacy} used EM to obfuscate both the structure and sensitive attributes of such networks. For datasets with mixed sensitivity, \cite{kamalaruban2020not} proposed selective privacy frameworks that prioritize protecting sensitive attributes during queries and synthetic data generation, achieving optimal trade-offs between privacy and utility through EM. The principles of EM have also been used to strengthen cryptographic primitives like Order-Preserving Encryption \cite{roy2022strengthening}.

While the theoretical adaptability of the EM is vast, a persistent practical challenge has been the computational cost of finding optimal or near-optimal mechanisms. Addressing this scalability issue is crucial for real-world deployment. The work by \cite{imola2022balancing} provides a significant step forward, proposing a framework that cleverly uses the EM's own formula as an \emph{equality constraint} within a larger Linear Program optimization. This hybrid approach dramatically reduces the complexity of the problem, making it feasible to compute high-utility mechanisms for large metric spaces. In a similar vein, \cite{zhang2022geo} introduced a \emph{Multi-Objective Evolutionary Algorithm (MOEA)} that uses EM as its core privacy engine but leverages the evolutionary algorithm to generate a set of Pareto-optimal privacy configurations for users to choose from, effectively managing the trade-off between multiple objectives like privacy and service quality.

\subsection{Optimization based Mechanism}
\label{subsec:optimization}


While the Laplace and EM effectively achieve mDP, they may not always provide the optimal balance between privacy and utility as they base noise probability on perturbation magnitude $d(x, y)$, which doesn’t fully capture the varied utility losses caused by perturbation in different directions. To tackle this issue, many recent efforts have increasingly focused on optimization-based approaches. 

Given the computational challenges of optimizing $\mathcal{M}$ when the secret data domain $\mathcal{X}$ and the perturbation data domain $\mathcal{Y}$ are continuous, a common strategy is to discretize both $\mathcal{X}$ and $\mathcal{Y}$ into finite sets \cite{imola2022balancing}. Under this framework, the random mechanism $\mathcal{M}$ is represented as a stochastic \emph{perturbation matrix} $\mathbf{Z} = \left[z_{x,y}\right]_{(x,y) \in \mathcal{X}\times \mathcal{Y}}$, where each entry $z_{x,y} = \Pr[\mathcal{M}(x) = y]$ denotes the probability of selecting $y \in \mathcal{Y}$ as the perturbed output for a given real input $x \in \mathcal{X}$. The perturbation matrix $\mathbf{Z}$ can then be optimized by solving the following \emph{linear programming (LP)} problem:
 \looseness = -1
\small
\begin{eqnarray}
\label{eq:optimal_mech}
\min && \sum_{s \in \mathcal{X}, y \in \mathcal{Y}} \pi_x z_{x, y} c_{x, y} \\
\text{s.t. } && z_{x, y} \leq e^{\epsilon d_{\mathcal{X}}(x, x')} z_{x', y}, \forall x, x' \in \mathcal{X}, y \in \mathcal{Y}, \\
    && \sum_{y \in \mathcal{Y}} z_{x, y} = 1, \forall x \in \mathcal{X}, \quad z_{x, y} \geq 0, \forall (x, y) \in \mathcal{X}\times \mathcal{Y}
\end{eqnarray}
\normalsize
Here, $c_{x, y}$ represents the utility loss caused by perturbed data $y$ given the real data $x$, and $\pi_x$ is the prior distribution of $x$ over the input domain $\mathcal{X}$.

The earliest optimization-based data perturbation mechanism dates back to 2014 when \cite{bordenabe2014optimal} first proposed this LP formulation to maximize utility while preserving Geo-Ind. However, a critical challenge identified in this seminal work is that the number of constraints in this LP is $O(|\mathcal{X}|^3)$, rendering it computationally infeasible for large domains \cite{imola2022balancing, AhujaEDBT2019}. This scalability issue has since become a central theme, motivating a significant line of research. The initial work by \cite{bordenabe2014optimal} suggested using spanner graphs to approximate the metric and reduce constraints. Subsequent works explored more advanced methods. For instance, \cite{AhujaEDBT2019} introduced a multi-step mechanism that applies optimization recursively on a hierarchical index. The framework in \cite{imola2022balancing} developed `ConstOPTMech`, a method that reduces the LP size by fixing far-away perturbation probabilities to a weighted exponential form and only optimizing over near neighbors. More recently, \cite{qiu2024enhancing} tackled the problem by partitioning the dataset and applying Benders Decomposition to solve the large-scale LP problem efficiently.

Beyond directly optimizing the perturbation matrix, another line of work focuses on enhancing the utility of existing, more efficient mechanisms through post-processing optimization. A notable contribution is the Bayesian remapping technique developed by \cite{chatzikokolakis2017efficient}, which uses prior knowledge to transform the output of a standard mechanism (like Planar Laplace) to a new point with higher expected utility, without compromising the original privacy guarantees. Other works have focused on optimizing the ``shape'' of the output, such as the ``geo-indistinguishable masking'' approach, which generates an optimal set of points (a mask) instead of a single point to better confuse adversaries \cite{lin2023geo}. Addressing utility from the user's perspective, \cite{Pappachan-EDBT2023} introduced the CORGI framework, which generates a robust obfuscation function on the server that users can then safely customize on their end (e.g., by pruning undesirable locations) without violating Geo-Ind properties.

The application of these optimization frameworks is particularly prominent in the field of Mobile and Spatial Crowdsourcing (MCS). \cite{wang2017location} developed a \emph{mixed integer nonlinear programming (MINLP)} model to optimize travel distances under differentially private geo-obfuscation, a framework that was later extended to incorporate distortion privacy and multi-task allocation in \cite{wang2019mobile}. Recognizing the multi-faceted nature of utility, \cite{qian2021optimal} explicitly incorporated a ``task completion quality'' score into the optimization objective. To adapt these mechanisms to real-world road networks, a series of works by \cite{qiu2020location} and \cite{Qiu-CIKM2020} redefined utility and privacy constraints based on shortest-path distance and introduced techniques like Dantzig-Wolfe decomposition for efficient solving. \cite{zhang2022task} also tackled this problem by proposing a group-based noise addition mechanism. For time-sensitive applications, \cite{qiu2024fine} proposed a fine-grained geo-obfuscation method that confines perturbations to peer location sets, thereby reducing the estimation error of travel distances under strict time constraints. A similar approach was used in \cite{zhang2018shiftroute} to protect route queries by shifting endpoints to nearby points-of-interest, with the selection probabilities determined by an LP problem.

As scenarios grew more complex, researchers began exploring alternative optimization paradigms. In cooperative localization, \cite{shi2021clap} employed \emph{Contract Theory} to design an optimal incentive mechanism under information asymmetry, while \cite{yu2023balancing} used a \emph{game-theoretic framework} to balance accuracy and privacy. For high-dimensional problems like task allocation in 3D space, where traditional methods face the curse of dimensionality, \cite{min2023geo} pioneered the use of \emph{deep reinforcement learning} (specifically, the A3C algorithm) to learn an optimal geo-perturbation and task allocation policy. In the context of IoT and smart homes, \cite{liu2018epic} used an LP to select proxy gateways to minimize energy consumption under DP constraints, while \cite{qiu2022trafficadaptor} used optimization to select an optimal ``fake trajectory'' from a candidate pool to defend against traffic flow-aware attacks.

Finally, some works have focused on the theoretical foundations and critiques of these optimization-based methods. The work in \cite{koufogiannis2015optimality} provided a theoretical underpinning by proving the optimality of the Laplace mechanism for identity queries via a dual-problem formulation. More abstractly, \cite{boedihardjo2024metric} offered a novel perspective by using tools from \emph{metric geometry}, revealing that the optimal privacy-utility tradeoff is intrinsically linked to the geometric properties of the underlying data space. The research community also performs critical self-correction; for example, \cite{shun2021differential} rigorously analyzed the PIVE framework from \cite{Yu-NDSS2017}, identified theoretical flaws in its privacy proofs, and proposed feasible correction schemes. These diverse approaches, from foundational LP models to advanced game-theoretic frameworks, collectively demonstrate a rich and evolving landscape for optimization-based privacy-preserving mechanisms.

\subsection{Comparison}
\label{subsec:comparison}

\begin{table*}[t]
\vspace{-0.00in}
\caption{Comparison of major related works (``PN'' means ``privacy notion''. ``EIE'' means ``expected inference error'', In the column ``Location set size'', ``---'' means ``There is no limit to the size').}
\vspace{-0.00in}
\label{Tb:mDP:utlilitycompare}
\centering
\footnotesize 
\begin{tabular}{l|c|c|c|c|l}
\toprule
\multicolumn{1}{ c  }{}
&\multicolumn{5}{ c }{Features}
\\
\cline{2-6}
\multicolumn{1}{ c|  }{Obfuscation methods}
&\multicolumn{1}{ |c| }{Lap.}
&\multicolumn{1}{ |c }{Exp.}
&\multicolumn{1}{ |c }{Opt.}
&\multicolumn{1}{ |c }{$K$}
&\multicolumn{1}{ |c }{Utility loss definition} 
\\ 
\hline
\hline
CCS 2012 \cite{Shokri-CCS2012} &  &  & \checkmark & 30 & Expected \& max distance between real and reported locations  \\ 
CCS 2013 \cite{Andres-CCS2013} & \checkmark &  &  & --- & Expected distance between real and reported locations \\ 
CCS 2014 \cite{bordenabe2014optimal} &  & & \checkmark & 50 & Expected distance between real and reported locations\\ 
PETS 2015 (Privacy game) \cite{Reza-PoPET2015} &  & & \checkmark & 300 & Expected utility cost (depends on applications)\\ 
PETS 2015 (Elastic metric) \cite{chatzikokolakis2015constructing} &  & \checkmark & & --- & Expected distance between real and reported locations \\ 
ICDM 2016 \cite{Wang-CIDM2016} &  &  & \checkmark & 57 & Expected residual standard error\\ 
WWW 2017 \cite{Wang-WWW2017} &  &  & \checkmark & 16 & Expected travel distance\\ 
NDSS 2017 \cite{Yu-NDSS2017} &  &  & \checkmark & 50 & Expected distance between real and reported locations \\ 
CCS 2017 \cite{oya2017back} &  &  & \checkmark & 25 & Average \& worst-case quality loss (depends on applications) \\ 
TIFS 2017 \cite{tong2017jointly} & \checkmark &  & & --- & Task Utility (Satisfaction Ratio) \\ 
PETS 2017 \cite{chatzikokolakis2017efficient} & \checkmark &  & \checkmark & --- & Expected quality loss (depends on applications)\\ 
{\rd  CSF 2018 \cite{alvim2018metric}} & \checkmark & &  & 30 & Expected distance between real and reported locations \\ 
{\rd  TIFS 2018 \cite{hua2017geo}} & \checkmark &  & & --- & Task Utility (Distortion/LBS Accuracy/Precision and Recall) \\ 
{\rd  ICDM 2019 \cite{feyisetan2019leveraging}} & \checkmark & &  & 100--300 & Author predictions/Accuracy \\ 
{\rd  EDBT 2019 \cite{AhujaEDBT2019}} & \checkmark & &  & 100--300 & Expected utility cost \\ 
TMC 2020 \cite{qiu2020location} &  & & \checkmark & 300--400 & Expected difference between real and estimated travel distance\\ 
{\rd  PETS 2020 \cite{kamalaruban2020not}} & \checkmark & \checkmark &  & 50 & Mean-squared error \\ 
{\rd  WSDM 2020 \cite{feyisetan2020privacy}} &  & \checkmark &  & 50--300 & Expected utility cost (depends on applications) \\ 
PETS 2020 \cite{Mendes-PETS2020} &  \checkmark & & & --- & Expected distance between real and reported locations \\ 
{\rd  ACL 2021 \cite{yue2021differential}} &  & \checkmark & & --- & Task Utility (Accuracy) \\ 
{\rd  TDSC 2021 \cite{wang2019mobile}} & \checkmark &  & & --- & Expected travel distance/Acceptance ratio \\ 
{\rd  CVPR 2021 \cite{chen2021perceptual}} &  \checkmark & & & --- & Task Utility (Reidentification ratio) \\ 
{\rd  ICDE 2021 \cite{da2021react}} &  \checkmark & & & --- & Task Utility (Precision/Recall) \\ 
{\rd  TWC 2022 \cite{min20213d}} & \checkmark & &  & --- & Average expected error between the perturbed location and the actual location \\ 
SIGSPATIAL 2022 \cite{qiu2022trafficadaptor} &  & & & 100 & Expected difference between real and estimated travel distance \\ 
UAI 2022 \cite{imola2022balancing} &  & \checkmark & \checkmark & 400 & Expected utility cost (depends on applications) \\ 
{\rd  IoTJ 2022 \cite{shi2021clap}} & \checkmark & & \checkmark & --- & Task Utility (type-k cooperative nodes’ utility) \\ 
{\rd  IoTJ 2022 \cite{Chen-IoTJ2022}} &  & & \checkmark & --- & Task Utility (MSE) \\ 
{\rd  CCS 2022 \cite{roy2022strengthening}} &  & \checkmark &  & --- & Task Utility (Accuracy of range queries) \\ 
{\rd  TKDE 2023 \cite{zhao2022geo}} &  & \checkmark &  & 300 & The difference between the obfuscated and original covariance matrices \\ 
{\rd  TMC 2023 \cite{ren2022distpreserv}} &  & \checkmark &  & --- & Expected distance between real and reported locations \\ 
EDBT 2023 \cite{Pappachan-EDBT2023} &  & & \checkmark & 70 & Expected distance between real and reported locations\\ 
{\rd  IJCAI 2024 \cite{qiu2024enhancing}} &  & & \checkmark & 1000 & Expected data quality loss \\ 
EDBT 2024 \cite{qiu2024fine} &  & & \checkmark & 300--400 & Expected distance between real and reported locations \\ 
PETS 2025 \cite{qiu2025time} &  &  \checkmark &  \checkmark & 1,600 & Expected difference between real and estimated travel distance \\
CCS 2025 \cite{Liu-CCS2025} & & & \checkmark & 5,000 & Expected difference between real and estimated travel distance \\
\bottomrule
\end{tabular}
\vspace{-0.05in}
\end{table*}

In comparing Laplace, Exponential, and optimization-based mechanisms, each exhibits distinct strengths and limitations in terms of \emph{adaptability}, \emph{utility preservation}, and \emph{time efficiency}. Table \ref{Tb:mDP:utlilitycompare} provides a comparative overview of major mDP-related works, organized by the underlying obfuscation method, the size of the location set (when applicable), and the definition of utility loss employed. Across studies, utility is quantified in diverse ways, from geometric distortions such as expected or maximum distance, to task-driven metrics including accuracy, satisfaction ratio, or reidentification risk. Collectively, the table highlights the trajectory of research progress: from early reliance on simple Laplace-based methods to more sophisticated hybrid and optimization-driven approaches, reflecting both the breadth of application scenarios and advances in scalability.

\paragraph{Adapatibility} Currently, the Laplace mechanism remains the most widely used data perturbation method in mDP due to its simplicity and time-efficient implementation, adding Laplace noise directly to data values. Among the {\rd  133} references collected summarized by our work\footnote{The details can be found on GitHub \url{https://github.com/Kerwinxxp/metricDP/tree/main}}, 69 focus specifically on Laplace-based data perturbation, while 28 and 33 focus on EM and optimization-based approaches, respectively. 
However, as highlighted by \cite{carvalho2023tem}, Laplace is less effective for structured or combinatorial outputs (e.g., graphs and word embeddings) because directly adding noise can compromise structural integrity and overlook data density in the metric space. For instance, word embeddings exist in a discrete finite field, making it unlikely for a noisy vector to correspond to a valid word. To address this, \cite{fernandes2018author,feyisetan2019leveraging} employed nearest neighbor approximations for the noisy vectors. These approaches treat the representation space as non-sensitive, which might overlook privacy loss during the nearest neighbor search. In such scenarios, EM \cite{carvalho2023tem} and optimization-based methods \cite{imola2022balancing} offer greater flexibility, especially for non-numeric outputs or selections from finite sets.

\paragraph{Utility-preserving} 
The Laplace mechanism adds noise without considering the region's density where the vector resides, potentially reducing the utility of data perturbation. In such cases, EM \cite{carvalho2023tem} provides greater flexibility and often delivers higher utility, particularly for non-numeric outputs or selections from finite sets. However, similar to the Laplace mechanism, EM selects perturbed data based solely on the perturbation magnitude $d(x, y)$, which does not fully account for the varied utility losses caused by perturbations in different directions. In contrast, optimization-based mechanisms can achieve lower data utility loss by explicitly considering the utility loss of each possible perturbation choice, including both magnitude and direction. This approach enables these mechanisms to minimize the expected utility loss, thereby achieving superior performance.


\paragraph{Time-efficiency and scalability} Both the Laplace mechanism and the EM algorithm are recognized for their scalability and relatively low computational complexity, making them practical for processing large datasets. Their typical computational complexity ranges between $O(1)$ and $O(n^2)$, where $n$ denotes the size of secret data domain (i.e., $|\mathcal{X}|=n$). For example, the Laplace mechanism proposed by \cite{Andres-CCS2013} and the EM algorithm introduced by \cite{chatzikokolakis2015constructing} demonstrate time complexities of $O(1)$ and $O(n)$, respectively, when used to protect location data in a 2-dimensional plane. Subsequent studies, such as \cite{mendes2018effect} and \cite{takagi2019geo}, reveal that in high-dimensional settings, both methods can experience higher time complexities, often escalating to $O(n^2)$ or beyond. To address these computational challenges, researchers commonly employ clustering techniques or greedy algorithms, effectively reducing the computational burden and enabling the Laplace mechanism and EM algorithm to scale efficiently for large datasets.  \looseness = -1

As opposed to Laplace and EM, optimization-based methods remain limited to small-scale secret data domains, although recent works have made efforts to improve scalability. For instance, \cite{AhujaEDBT2019} proposed a multi-step algorithm that protects location data under Geo-Ind by leveraging a hierarchical index structure. Similarly, \cite{imola2022balancing} proposed a hybrid approach that applies EM to data with minimal utility impact and LP to data with greater utility sensitivity, improving computational efficiency while preserving privacy guarantees in large metric spaces. 
Another line of work, including \cite{qiu2020location,qiu2024enhancing,qiu2025time,Liu-CCS2025}, employed optimization decomposition techniques such as Dantzig–Wolfe decomposition and Benders decomposition to break down large-scale LP problems into smaller, manageable subproblems, thereby improving the scalability of these solutions. 

\vspace{-0.1in}
\section{Applications} 
\label{sec:applications}

Since its introduction in 2013, mDP has proven to be a versatile and effective framework across diverse application domains, largely due to its ability to tailor privacy guarantees to domain-specific distance metrics. As shown in Figure \ref{fig:mDPhistory}, research activity has grown steadily, with location-based applications remaining the dominant focus, while increasing interest in text, image, and voice data highlights mDP’s expanding reach and influence.

\vspace{-0.03in}
\subsection{Application Domains} 
\label{subsec:applications}

\vspace{-0.00in}


\paragraph{Location privacy} To date, the most extensively studied application domain of mDP is location privacy protection. The application of mDP in this area can be traced back to 2013, when Andr{'e}s \emph{et al.} \cite{Andres-CCS2013} first extended differential privacy to geographic data (Geo-Ind) and proposed the polar Laplace mechanism to perturb users’ locations in a two-dimensional space. Building on this foundation, a substantial body of research has further advanced location privacy techniques using Laplace, EM, and optimization-based approaches. These efforts address diverse privacy and utility requirements across various LBS applications, including \emph{mobile spatial crowdsourcing (MSC)} and \emph{intelligent transportation systems (ITS)}.

For instance, in MSC, mDP has been instrumental in addressing task allocation challenges while preserving worker location privacy. Wang \emph{et al.} \cite{wang2017location} proposed a differential geo-obfuscation mechanism that balances privacy and utility by optimizing workers’ travel distances for task assignments. Zhang \emph{et al.} \cite{zhang2022area} developed a method that achieves a trade-off between location privacy and efficient area coverage. This approach was later extended by Ma et al. \cite{ma2022personalized}, who introduced a personalized noise addition framework, enabling privacy guarantees tailored to individual users' preferences. Zhang \emph{et al.}  \cite{zhang2022geo} further integrated the exponential mechanism with evolutionary algorithms to enhance task allocation under privacy constraints. Recent advancements also include reinforcement learning-based techniques by Min \emph{et al.}  \cite{min2023geo} and scalable optimization methods proposed by Qiu \emph{et al.} \cite{qiu2020location}, which are designed to handle large-scale crowdsourcing environments more effectively.

To protect vehicles’ location privacy, Takagi \emph{et al.} \cite{takagi2019geo} applied road network metrics to enhance obfuscation in constrained spatial environments. They extended Geo-Ind to road networks by proposing \emph{Geo-Graph-Indistinguishability}, which replaces the Euclidean distance with shortest-path metrics to account for vehicles’ mobility constraints on road networks. These methods, when combined with dynamic, context-aware adaptations, such as those proposed by Yu \emph{et al.} \cite{Yu-NDSS2017}, can provide real-time privacy guarantees tailored to user preferences. Li \emph{et al.} \cite{li2020perturbation} introduced the “Perturbation-Hidden” mechanism, which employs the exponential mechanism to inject noise into vehicle location data. This approach strikes a balance between privacy and the accuracy demands of LBS in the Internet of Vehicles, effectively mitigating tracking and inference attacks while preserving utility.

The adaptability of mDP is further exemplified by its integration with emerging technologies. For instance, Yang \emph{et al.} \cite{yang2021blockchain} combined a decentralized blockchain framework with a modified geo-indistinguishability mechanism, termed \emph{Truncated Geo-Indistinguishability (TGeoI)}. In the context of an indoor paging service, this approach perturbs user locations within a predefined radius, effectively preventing the generation of invalid locations (e.g., outside a building) while still offering provable privacy guarantees. This integration highlights the potential of mDP to support secure data sharing in environments involving untrusted parties.


\paragraph{Text data privacy} The application of mDP has also been explored in the domain of text perturbation. Fernandes \emph{et al.} \cite{fernandes2018author} applied the Laplace mechanism to the Word Mover Distance to anonymize author identities while preserving semantic integrity. Similarly, Arnold \emph{et al.} \cite{arnold2023guiding} advanced text privatization by introducing syntactically guided Laplace noise, ensuring semantic coherence alongside privacy and extending the applicability of differential privacy in natural language processing.

However, Laplace mechanisms are less effective for perturbing word embeddings, as directly adding noise to vectors in a discrete finite vocabulary often results in invalid or semantically irrelevant words. To address this challenge, Fernandes \emph{et al.} \cite{fernandes2018author} and Feyisetan \emph{et al.} \cite{feyisetan2019leveraging} proposed using nearest neighbor approximations for noisy vectors, treating the embedding space as non-sensitive to avoid additional privacy leakage during nearest neighbor searches. Despite these improvements, such methods often overlook the varying density of the embedding space, which can lead to suboptimal utility. \looseness = -1

In contrast, EM offers greater flexibility and utility, especially for non-numeric outputs or selections from finite sets. For instance, Wang \emph{et al.} \cite{wang2017local} employed a structured EM to estimate ordinal data distributions, leveraging inherent order to reduce estimation error while preserving privacy. Feyisetan \emph{et al.} \cite{feyisetan2020privacy} further applied EM to perturb text embeddings, maintaining semantic coherence without compromising privacy. Yue \emph{et al.} \cite{yue2021differential} proposed mechanisms to generate sanitized text tailored to downstream tasks such as sentiment analysis and classification.

Recent advancements include Carvalho \emph{et al.} \cite{carvalho2023tem}’s \emph{Truncated Exponential Mechanism (TEM)}, which adapts noise based on local embedding density, and Meisenbacher \emph{et al.} \cite{meisenbacher20241}’s 1-Diffractor, which simplifies computations by projecting embeddings to a single dimension while preserving differential privacy guarantees.

\paragraph{Image data privacy} Image data has seen a parallel evolution in privacy protection techniques. An early approach by Fan \emph{et al.} \cite{fan2019practical} applied mDP to the \emph{Singular Value Decomposition (SVD)} feature space of images, perturbing singular values to achieve provable privacy guarantees. Building on the idea of perturbing in a feature space, Chen \emph{et al.} \cite{chen2021perceptual} proposed PI-Net, a framework that defines ``\emph{Perceptual Indistinguishability}'' in the latent space of a \emph{Generative Adversarial Network (GAN)}. By applying a Laplace-based mechanism to these semantically meaningful latent codes, PI-Net can obfuscate facial images while preserving high-level visual attributes.

To further improve the utility of protected images, Yan \emph{et al.} \cite{yan2024coder} introduced CODER, a mechanism that incorporates a composite metric combining the traditional Euclidean ($\ell_2$) distance with Jensen-Shannon (JS) divergence. This hybrid metric captures both geometric and distributional differences in pixel histograms, thereby enhancing the preservation of structural information in the perturbed images.

\vspace{-0.00in}
\paragraph{Other applications} More recently, mDP has been applied beyond spatial, textual, and image data, extending to a variety of domains including images, voice, graph data, and complex distributed systems.

In the realm of biometric data, Han \emph{et al.} \cite{han2020voice} introduced the notion of ``\emph{voice-indistinguishability}'' by extending mDP to protect voiceprints. Their method represents voiceprints as high-dimensional x-vectors and adds noise calibrated to the angular distance between them, ensuring that similar-sounding voices receive stronger privacy guarantees against spoofing and identification attacks.

Beyond perceptual media (e.g., images and voice data), mDP has been applied to structured domains by choosing an application-specific metric, including graph-structured data and network telemetry. A prominent example is node-differential privacy, where the distance metric is defined by the number of node additions or deletions required to transform one graph into another, an application that naturally aligns with the principles of mDP \cite{borgs2018revealing, raskhodnikova2016lipschitz}.

Raskhodnikova \emph{et al.} \cite{raskhodnikova2016lipschitz} laid the groundwork for this area by introducing key techniques, including efficiently computable Lipschitz extensions for vector-valued graph statistics (e.g., degree distributions) and a generalized exponential mechanism. Building upon these tools, Borgs \emph{et al.} \cite{borgs2018revealing} developed node-private algorithms for estimating parameters of complex network models such as graphons, demonstrating how to uncover macroscopic network structure while preserving the privacy of individual nodes.

Shifting from static datasets to complex interactive systems, the principles of mDP have been extended to domains such as federated learning and the Internet of Things (IoT). In distributed machine learning, Galli \emph{et al.} \cite{galli2023advancing} applied mDP, referred to as $d$-privacy, to provide group-level privacy in Personalized Federated Learning (PFL). By injecting Laplace noise into model parameter updates, their framework ensures that clients with similar data distributions form a privacy-protected group, achieving a balance between model personalization and formal privacy guarantees.

In the context of IoT, Liu \emph{et al.} \cite{liu2018epic} proposed the EPIC framework to defend smart homes against internet traffic analysis. It employs an optimization-based differential privacy mechanism to select proxy gateways, effectively applying mDP to the physical locations of smart homes. This approach obfuscates the traffic source, preventing adversaries from linking observed activity to specific households.

\begin{table*}[ht]
\centering
\caption{Well-known datasets used in mDP-related research.}
\label{tab:mdp-datasets-cleaned-cited}
\begin{tabular}{p{2.6cm}|p{2.1cm}|p{4.5cm}|p{2.8cm}|p{2.4cm}}
\toprule
\textbf{Dataset Name} & \textbf{Type} & \textbf{Application Domain} & \textbf{Size} & \textbf{Citing References} \\
\hline
\hline
GeoLife \cite{zheng2010geolife} & Location & Trajectory obfuscation, semantic-aware perturbation, adaptive privacy modeling & 17,621 trajectories from 182 users &\cite{chatzikokolakis2014predictive, al2018adaptive, ma2018agent, mendes2018effect, hua2017geo, mendes2020impact, cao2020pglp} \\
\hline
T-Drive \cite{yuan2010t} & Location & Vehicle routing, online task assignment, frequent query perturbation, trajectory sanitization & Trajectories of 10,357 taxis (~15M points) & \cite{chatzikokolakis2014predictive, tong2017jointly, hua2017geo, to2018privacy, li2023privacy} \\
\hline
Gowalla/Brightkite \cite{cho2011gowalla} & Location/Social & LBSN privacy, utility metrics, contact graph inference, spatial crowdsourcing & G: 196k users, 6.4M check-ins; B: 58k users, 4.5M check-ins & \cite{chatzikokolakis2015constructing, chatzikokolakis2017efficient, oya2017geo, to2018privacy, da2021react, cao2020pglp, li2023accurate} \\
\hline
Foursquare \cite{yang2013foursquare} & Location/Check-in & POI recommendation, mobility modeling, friend matching under LDP & NYC: 227k check-ins; Tokyo: 573k check-ins & \cite{wang2018personalized, kawamoto2019local, fernandes2021locality} \\
\hline
Cabspotting / Portocabs \cite{piorkowski2009cabspotting, moreira2013portocabs} & Location & Urban mobility modeling, map-matching attack evaluation, location frequency analysis & Cabspotting: ~500 taxis (30 days); Portocabs: 441 taxis (1 year) & \cite{mendes2020impact} \\
\hline
MovieLens \cite{harper2015movielens} & Ratings & Collaborative filtering, friend matching under embedding perturbation & Varies (e.g., 25M version: 162k users, 25M ratings) & \cite{fernandes2021locality} \\
\hline
Yelp \cite{yelp_dataset_challenge} & Location/Text & Location-aware crowdsensing, privacy-incentive tradeoff modeling & Millions of reviews, users, businesses (periodically updated) & \cite{huang2019incentivizing} \\
\hline
EMNIST \cite{cohen2017emnist} & Image & Federated learning under group privacy, fairness-aware digit classification & 814,255 total characters from 3,500 writers (62 classes) & \cite{galli2023group} \\
\hline
IMDb Dataset \cite{maas2011imdb} & Text/Corpus & Text-to-text privatization, sentiment masking for LLMs & 50k labeled reviews + 50k unlabeled docs & \cite{arnold2023guiding} \\
\hline
PAN11 / PAN12 (Enron) \cite{potthast2011pan, stamatatos2012pan} & Text/Email & Author masking, stylometric obfuscation, attribution perturbation & Corpora of varying sizes per task/year & \cite{feyisetan2019leveraging, Fernandes-PST2019} \\
\hline
NLP Benchmarks \cite{pang2005seeing, hu2004mining} & Text & Embedding-level privatization, sentiment masking, LLM fine-tuning & Varies (e.g., MR: 10.7k, CR: 3.8k samples) & \cite{feyisetan2019leveraging, feyisetan2021private} \\
\hline
US Cities Database \cite{simplemaps_uscities} & Geospatial/Census & Histogram queries over spatial bins, dx-private perturbation & Free version: >40k U.S. cities & \cite{kamalaruban2020not} \\
\hline
US Census/ACS \cite{uscensusbureau_acs} & Tabular/Census & Population histogram release, policy-aware DP, tabular domain modeling & Millions of Public Use Microdata Samples (PUMS) & \cite{haney2014design} \\
\hline
Twitter Geo-location \cite{twitter_api} & Location/Count & Social media aggregation, spatial DP for geo-located activity & Streaming API; size depends on collection parameters & \cite{haney2014design} \\
\hline
Network/Citation Graphs \cite{yang2016revisiting} & Graph/Temporal & Graph DP, node classification, edge perturbation & Varies (e.g., Cora: 2.7k nodes, 5.4k edges) & \cite{haney2014design} \\
\hline
OpenStreetMap \cite{openstreetmap_foundation} & Geospatial/Map & POI indexing, road network perturbation, urban topology analysis & Global, community-updated road network and POIs & \cite{chatzikokolakis2015constructing, chatzikokolakis2017efficient, hua2017geo} \\
\hline
GRUMPv1 \cite{balk2004grump} & Geospatial/Census & Population grid modeling, density-aware spatial privacy & Global grid (~1km or 30 arc-second resolution) & \cite{koufogiannis2016location} \\
\hline
Google Places API \cite{google_places_api} & Location/POI & POI recommendation, geo-query sanitization, location-based tasks & Query-dependent real-time data & \cite{eltarjaman2017location} \\
\hline
Synthetic Datasets & Location & Controlled mDP benchmarks: LBS simulation, indoor navigation, grid cities & Task-dependent (user-defined parameters) & \cite{elsalamouny2014generalized, Koufogiannis-JPC2017, Chatzikokolakis-PETS2013, huang2016eppd, elsalamouny2016differential, qiu2024fine, qiu2024enhancing, kamalaruban2020not, min20213d} \\
\bottomrule
\end{tabular}
\vspace{-0.1in}
\end{table*}


\subsection{Context-Aware mDP Mechanisms}
\label{subsec:context}
Notably, mDP has evolved through applications that dynamically adjust privacy protections based on contextual information or user behavior. Particularly, \cite{chatzikokolakis2014predictive} first introduced a predictive mDP mechanism that adjusts noise levels according to historical mobility patterns, enhancing privacy protection in mobility trace data. After that, \cite{haney2014design} and \cite{Reza-PoPET2015} explored policy-driven and game-theoretic approaches to privacy settings, enabling customized configurations aligned with user-specific policies or threat models. \cite{Yu-NDSS2017} proposed a mechanism to adjust noise levels based on user preferences, offering flexible privacy guarantees in location-based services (LBS).
\cite{eltarjaman2017location} proposed a rank-based approach that modulates privacy protections based on query importance in geo-query systems. This method balances privacy and utility by prioritizing more accurate responses for top-ranked queries. 

More recently, \cite{cao2020pglp} proposed PGLP, a framework allowing users to define privacy settings based on their location history or preferences, ensuring privacy tailored to individual needs. \cite{zhao2022geo} introduced Geo-Ellipse Indistinguishability, incorporating community-level covariance matrices to provide privacy adapted to data dispersion and orientation in spatial datasets. Furthermore, \cite{Pappachan-EDBT2023} developed a customizable geo-obfuscation framework using a tree-structured organization of locations at varying granularity levels, enabling privacy based on diverse user demands. \cite{odoh2024group} presented a group-wise k-anonymity scheme combined with differential privacy, dynamically considering group characteristics to enhance protections against inference attacks and provide robust privacy guarantees.

\subsection{Dataset}

The evaluation and advancement of mDP mechanisms rely heavily on a diverse range of public and synthetic datasets that serve as benchmarks for assessing privacy-utility trade-offs. The selection of a dataset is often guided by the specific application domain being addressed, with a significant majority of research focusing on location-based scenarios, followed by growing interest in text, image, and other data types. Table~\ref{tab:mdp-datasets-cleaned-cited} provides a comprehensive summary of datasets commonly used in mDP-related research.

Consistent with the application trends shown in Figure~1, location and mobility datasets are the most widely used benchmarks in mDP literature. Trajectory datasets, such as GeoLife~\cite{zheng2010geolife} and T-Drive~\cite{yuan2010t}, are fundamental for developing and testing mechanisms for continuous location privacy. GeoLife, containing over 17,000 trajectories from 182 users, is frequently employed to evaluate adaptive and semantic-aware perturbation techniques. For instance, it has been used to assess predictive mechanisms~\cite{chatzikokolakis2014predictive}, policy-driven privacy frameworks~\cite{cao2020pglp}, and the impact of update frequency on privacy~\cite{mendes2018effect, mendes2020impact}. The larger T-Drive dataset, with trajectories from over 10,000 taxis, is often used to model scenarios with denser data, such as privacy-preserving vehicle routing, online task assignment in spatial crowdsourcing, and trajectory sanitization~\cite{chatzikokolakis2014predictive, tong2017jointly, hua2017geo, to2018privacy, li2023privacy}.

\emph{Location-based social network (LBSN)} check-in data, including the Gowalla and Brightkite datasets~\cite{cho2011gowalla}, provide a rich intersection of spatial and social graph information. These datasets are instrumental in studies on LBSN privacy, the development of utility metrics, and analyzing risks like contact graph inference~\cite{chatzikokolakis2015constructing, chatzikokolakis2017efficient, oya2017geo, to2018privacy, da2021react, cao2020pglp, li2023accurate}. Similarly, the Foursquare dataset~\cite{yang2013foursquare} is a popular benchmark for developing privacy-preserving point-of-interest (POI) recommendations and mobility models~\cite{wang2018personalized, kawamoto2019local, fernandes2021locality}. For urban mobility and transportation systems, datasets like Cabspotting and Portocabs~\cite{piorkowski2009cabspotting, moreira2013portocabs} offer real-world vehicle GPS data used to evaluate location frequency analysis and defenses against map-matching attacks~\cite{mendes2020impact}.

Beyond location data, mDP has been increasingly applied to textual data. The IMDb Dataset~\cite{maas2011imdb} is used for developing text-to-text privatization and sentiment masking for large language models~\cite{arnold2023guiding}. For tasks focused on author privacy, the PAN11/PAN12 (Enron) email corpora~\cite{potthast2011pan, stamatatos2012pan} are standard benchmarks for stylometric obfuscation and attribution perturbation~\cite{feyisetan2019leveraging, Fernandes-PST2019}. Broader NLP Benchmarks, such as movie review datasets~\cite{pang2005seeing, hu2004mining}, are used to test embedding-level privatization and fine-tuning under mDP constraints~\cite{feyisetan2019leveraging, feyisetan2021private}.

Other data modalities are also emerging in mDP research. In the image domain, the EMNIST dataset~\cite{cohen2017emnist} has been used to explore fairness and group privacy in federated learning contexts~\cite{galli2023group}. Tabular and geospatial census data, such as the US Census/ACS~\cite{uscensusbureau_acs} and the US Cities Database~\cite{simplemaps_uscities}, are used to design policy-aware mechanisms and dx-private algorithms for releasing population histograms~\cite{haney2014design, kamalaruban2020not}. For research involving road networks, OpenStreetMap~\cite{openstreetmap_foundation} is a critical resource for POI indexing and network-aware perturbation~\cite{chatzikokolakis2015constructing, chatzikokolakis2017efficient, hua2017geo}, while the Google Places API~\cite{google_places_api} facilitates experiments on real-time geo-query sanitization~\cite{eltarjaman2017location}.

Finally, Synthetic Datasets play a crucial role in the theoretical and practical validation of mDP mechanisms. They provide controlled environments for benchmarking, allowing researchers to systematically evaluate algorithms under specific parameters, such as for indoor navigation simulations or grid-based location services, without the noise and irregularities of real-world data~\cite{elsalamouny2014generalized, Koufogiannis-JPC2017, Chatzikokolakis-PETS2013, huang2016eppd, elsalamouny2016differential, qiu2024fine, qiu2024enhancing, kamalaruban2020not, min20213d}.

\vspace{-0.15in}
\section{Future Prospects}
\label{sec:future}
The field of mDP has made significant progress over the past decade, but there are still many opportunities for advancement. Below, we outline key future directions for further development. 

\vspace{-0.00in}
\paragraph{Scalability and practical deployment} Scalability remains a major challenge for utility-preserving mDP, particularly with increasingly large and high-dimensional datasets. A promising future direction is to develop more efficient optimization methods that reduce computational overhead and can handle real-world, large-scale deployments. Simplified models that approximate mDP while maintaining acceptable privacy guarantees could also improve practicality for industrial applications. Techniques such as optimization decomposition have been explored to enhance scalability \cite{qiu2024enhancing,qiu2025time}, but further advancements are needed to make these methods more widely applicable.

\vspace{-0.00in}
\paragraph{Utility-preserving text data protection} While mDP research has traditionally focused on location data, there is a growing body of work exploring its application for protecting textual data to address privacy concerns in Large Language Models (LLMs). However, challenges remain in balancing privacy and utility, as excessive noise can degrade the quality of generated text, especially in nuanced contexts. The high dimensionality of LLMs and the complex dependencies between parameters complicate the direct application of mDP, requiring new techniques like adaptive noise mechanisms or task-specific privacy optimizations. Ensuring scalability for large-scale training and addressing privacy guarantees for compositional outputs are also critical hurdles. Furthermore, the rapid advancement of deep neural networks highlights limitations in the traditional mDP framework, which was designed primarily for statistical models, making it potentially insufficient to protect users' data against sophisticated attacks. This necessitates the development of new threat models and designing advanced countermeasures tailored to the unique vulnerabilities and complexities of deep learning systems, ensuring effective privacy protections in evolving applications.

\vspace{-0.00in}
\paragraph{Adaptive context-aware mechanisms} The development of adaptive, context-aware mechanisms is a promising direction for mDP. Future research should explore how privacy levels can be dynamically adjusted based on real-time contextual information, such as user behavior or data sensitivity. This would involve leveraging AI techniques to intelligently modulate privacy levels, ensuring a better balance between utility and privacy in diverse scenarios. Context-aware approaches like Geo-Ellipse-Indistinguishability \cite{zhao2022geo} have demonstrated the benefits of adjusting privacy parameters based on spatial contexts, and similar techniques could be expanded to other domains.

\vspace{-0.00in}
\paragraph{Enhancing theoretical foundations.} Further theoretical advancements are needed to deepen the understanding of mDP, particularly in the trade-offs between privacy, utility, and efficiency. Research could explore new types of distance metrics and theoretical models that better capture real-world data relationships. Additionally, establishing tighter bounds on privacy guarantees can help guide the practical implementation of mDP in various applications. Existing works like \cite{boedihardjo2024metric} have explored new metrics for privacy-utility trade-offs, but further development is needed to establish comprehensive theoretical frameworks.  

\vspace{-0.02in}
\section{Conclusion}
\label{sec:conclusions}

Over the past decade, mDP has evolved into a crucial extension of traditional DP, offering flexible, distance-based privacy guarantees that address the diverse requirements of applications such as location-based services, text processing, and multimedia privacy. This overview article has highlighted the development of key mDP mechanisms, including the Laplace mechanism, Exponential mechanism, and optimization-based approaches, as well as their applications and scalability. 

While significant progress has been made, challenges persist, particularly in scaling mDP for large datasets, preserving data utility across various domains, and dynamically adapting privacy protections based on context. Addressing these challenges through future research in efficient optimization, adaptive context-aware mechanisms, and deeper theoretical models will be essential to fully realize mDP’s potential in protecting privacy in emerging fields like large-scale machine learning, real-time systems, and personalized data protection.

\vspace{-0.0in}
\begin{IEEEbiography}[{\includegraphics[width=1.3in,height=1.3in,clip,keepaspectratio]{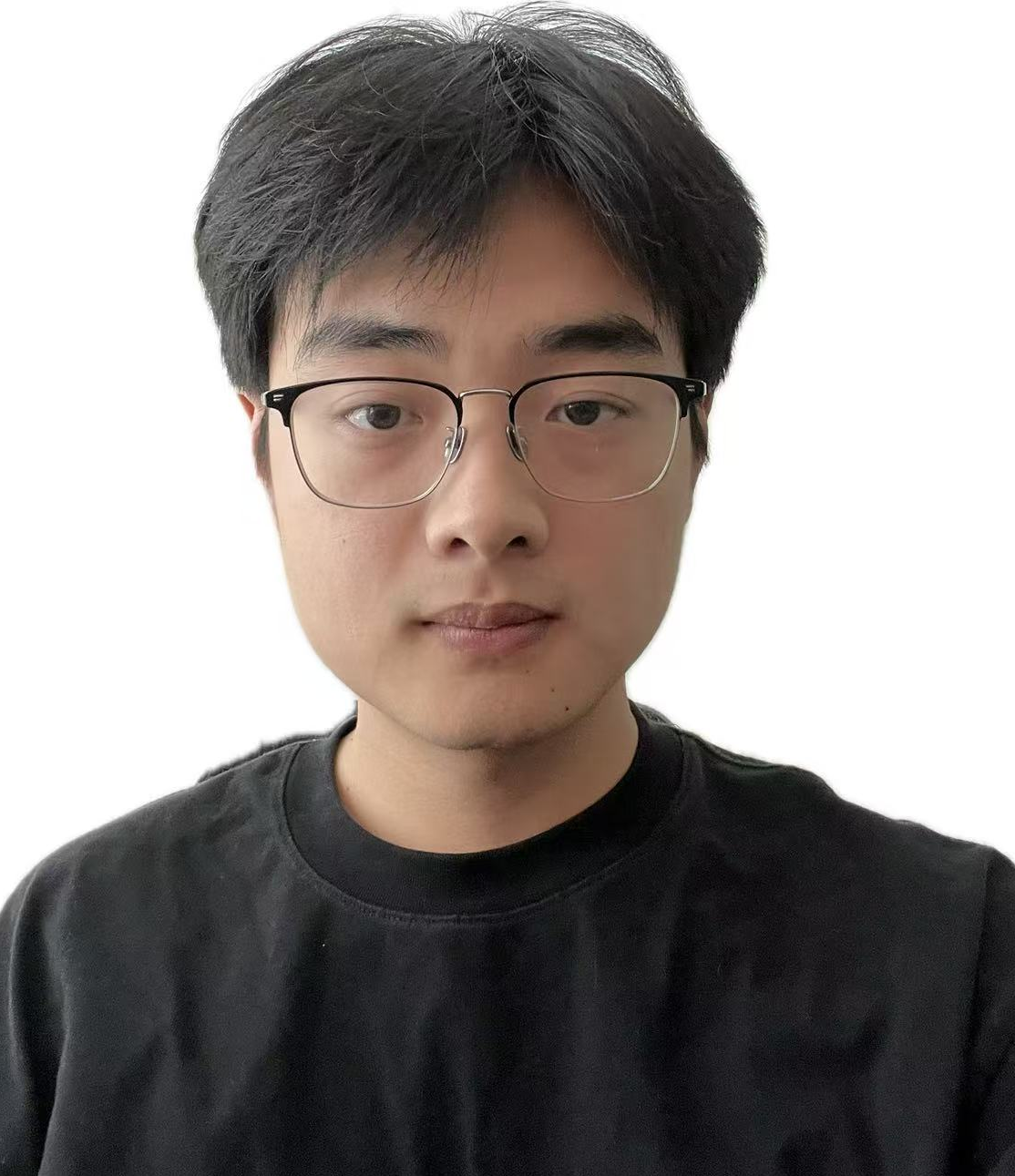}}]{Xinpeng Xie}
received his B.S. degree in Automation from Sichuan University, Chengdu, China, in 2023. He is currently pursuing the Ph.D. degree in Computer Science at the Department of Computer Science and Engineering, University of North Texas (UNT), Denton, TX. His research interests include spatial computing and data privacy protection.
\end{IEEEbiography}

\vspace{-0.0in}
\begin{IEEEbiography}[{\includegraphics[width=1.3in,height=1.3in,clip,keepaspectratio]{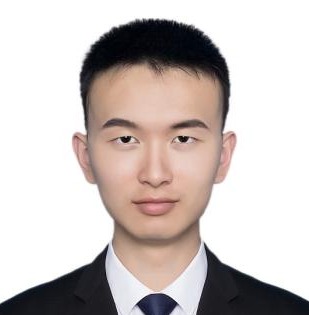}}]{Chenyang Yu} received his B.S. degree in Telecommunication Engineering from Wuhan University of Technology, Wuhan, China, in 2020. He is currently pursuing the Ph.D. degree in Computer Science at the Department of Computer Science and Engineering, University of North Texas (UNT), Denton, TX. His research interests include time series forecasting and large language models.
\end{IEEEbiography}

\begin{IEEEbiography}[{\includegraphics[width=1.3in,height=1.3 in,clip,keepaspectratio]
{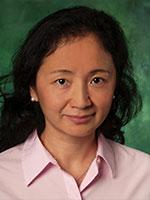}}]{Yan Huang} received her B.S. degree in Computer Science from Peking University, Beijing, China, in 1997 and a Ph.D. degree in Computer Science from the University of Minnesota, in 2003. She is a Regents Professor of the Department of Computer Science and Engineering at the University of North Texas. Her general research areas are geospatial artificial intelligence. Her contributions are recognized with the ACM SIGSPATIAL 10-year Impact Award (2019, ACM SIGSPATIAL 10-years Impact Award Runner up (2020). She is a recipient of the University of North Texas Decker Scholar award and is recognized as an ACM Distinguished Member for outstanding scientific contribution to computing in 2019. She has been awarded \$9 million in total by agencies including ARL, NSF, NGA, ONR, DOT, and state agencies. 

In professional societies, Dr. Huang has served in leadership roles as board member, General Chair, Program Committee Chair for flagship organizations such as ACM SIGSPATIAL and SSTD Endowment.  She served as the Financial Chair for IEEE International Conference on Data Engineering (ICDE 2020) and IEEE International Conference on Data Mining series (ICDM 2013). She was General Co-chair of ACMSIGSpatial Conference (2014 and 2015). She served as Program Committee Co-Chair for ACM SIGSpatial Conference (2020 and 2021). She was on Board of Director of SSTD Endowment, 2014-201 and 10-year Impact Paper Review Committee of ACM SIGSpatial Conference, 2023 and 2025.
\end{IEEEbiography}

\vspace{-0.0in}
\begin{IEEEbiography}[{\includegraphics[width=1.15in, height=1.15in,keepaspectratio]{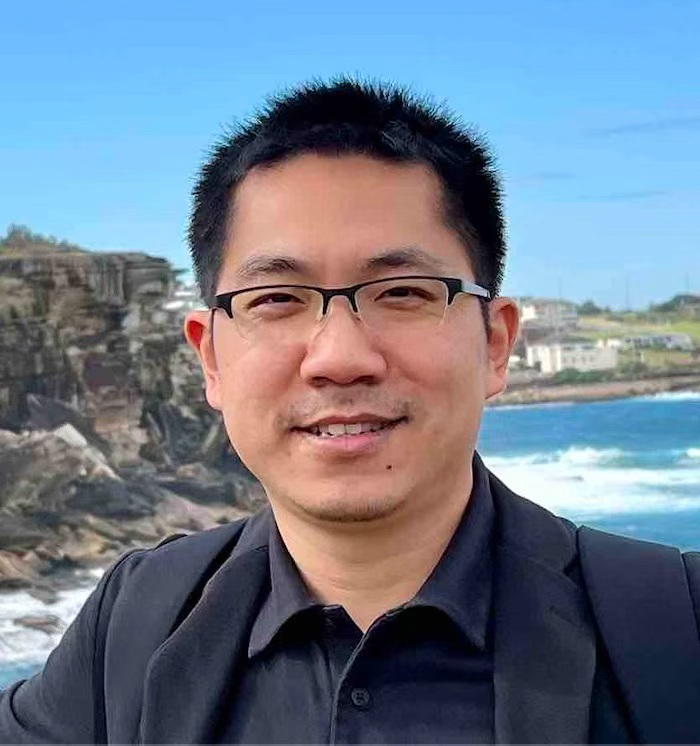}}]{Yang Cao} is an Associate Professor at the Department of Computer Science, Institute of Science Tokyo (Science Tokyo, formerly Tokyo Tech), and directing the Trustworthy Data Science and AI (TDSAI) Lab. He is passionate about studying and teaching on algorithmic trustworthiness in data science and AI. Two of his papers on data privacy were selected as best paper finalists in top-tier conferences IEEE ICDE 2017 and ICME 2020. He was a recipient of the IEEE Computer Society Japan Chapter Young Author Award 2019, Database Society of Japan Kambayashi Young Researcher Award 2021. His research projects were/are supported by JSPS, JST, MSRA, KDDI, LINE, WeBank, etc.
\end{IEEEbiography}

\begin{IEEEbiography}[{\includegraphics[width=1.3in,height=1.3 in,clip,keepaspectratio]
{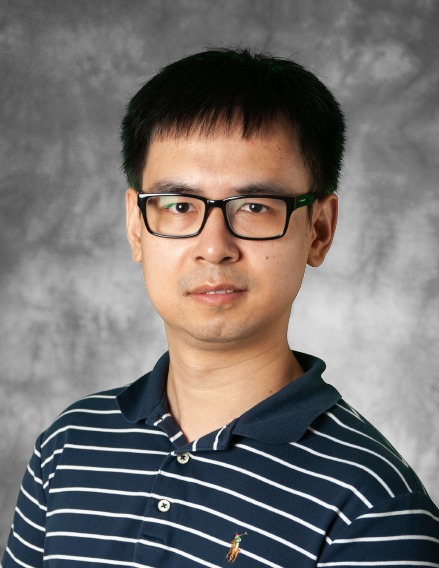}}]{Chenxi Qiu} (Member, IEEE) received the B.S. degree in Telecommunication Engineering from Xidian University, China, in 2009, and the Ph.D. degree in Electrical and Computer Engineering from Clemson University, USA, in 2015. He was a Postdoctoral Scholar at Penn State University from 2016 to 2018 and an Assistant Professor at Rowan University from 2018 to 2021. He is currently an Assistant Professor in the Department of Computer Science and Engineering at the University of North Texas. His research interests include statistical privacy and privacy-preserving computing. He received the Best Transactions Paper Award from the IEEE Transactions on Intelligent Transportation Systems and the Best Paper Award at IEEE CloudCom 2016.
\end{IEEEbiography}
\vspace{-0.0in}

\DEL{
\clearpage 
\appendix
\section{Laplace Mechanism}
The Laplace mechanism is among the most widely used data perturbation techniques, originally developed to achieve differential privacy (DP) \cite{Dwork-TC2006}. It perturbs the output of a query function $f(x)$ by adding noise $x$, resulting in:
\[
\mathcal{M}(f(x)) = f(x) + x,
\]
where $x$ is drawn from a Laplace distribution with a probability density function:
\[
\mathrm{Lap}(x) = \frac{\epsilon}{2\Delta f} e^{-\frac{\epsilon |x|}{\Delta f}}.
\]
Here, $\Delta f = \sup_{x, y \in \mathcal{X}} |f(x) - f(x')|$ represents the global sensitivity of $f$, measuring the maximum possible change in the function's output between neighboring datasets.

The Laplace mechanism naturally adapts to various metrics by appropriately scaling the noise \cite{McSherry-FOCS2007}. To extend it for metric differential privacy (mDP), \cite{Chatzikokolakis-PETS2013} proposed scaling the noise based on a metric $d_{\mathcal{X}}$ instead of the global sensitivity $\Delta f$. This metric, $d_{\mathcal{X}}(s, x)$, quantifies the distinguishability between $x$ and $x$. The resulting probability density function is expressed as:
\[
\mathrm{Lap}(x) = \lambda(x) e^{-d_{\mathcal{X}}(s, x)}.
\]
Building on this, \cite{Andres-CCS2013} introduced polar Laplace noise to achieve \emph{geo-indistinguishability (Geo-Ind)}. This approach employs a two-dimensional Laplace distribution centered at the true location $x$, with a probability function given by:
\[
\Pr\{\mathcal{M}(x) = y\} = \frac{\epsilon^2}{2\pi} e^{-\epsilon d(x, y)}.
\]

Following \cite{Andres-CCS2013}, numerous studies have applied the Laplace mechanism to achieve Geo-Ind. For example, \cite{chatzikokolakis2014predictive} tailored Laplace noise for synthetic route generation in smartphone applications, while \cite{wang2014entropy} proposed an entropy-minimizing mechanism that combines Laplace noise with system feedback control. Other notable works include \cite{arcolezi2021preserving}, which adapted Geo-Ind for emergency location sharing, and \cite{shi2021clap}, which tailored it for incentivized crowdsensing platforms. \cite{min20213d} extended Geo-Ind to three-dimensional indoor spaces to address challenges in high-resolution spatial data. Similarly, \cite{min2023personalized} developed personalized 3D location privacy techniques using distortion-based geo-perturbation. \cite{yang2021blockchain} integrated blockchain technology with Laplace noise for secure indoor data anonymization.

Beyond basic Geo-Ind, advanced use cases have emerged in trajectory and mobility trace privacy. For instance, \cite{chatzikokolakis2014predictive} generalized Geo-Ind to mobility trace privacy by leveraging predictive correlations across multiple location reports. \cite{mendes2018effect} examined update frequencies and proposed adaptive noise calibration to balance privacy and utility.

Protecting trajectory data became particularly critical during the COVID-19 pandemic. \cite{li2023accurate} proposed a Laplace-based mechanism for accurate contact tracing, while \cite{yu2023privacy} developed longitudinal Geo-Ind to protect trajectory privacy in advertising systems. Other studies, such as \cite{li2023privacy} and \cite{duarte2024privacy}, refined these approaches by employing tailored metrics and adaptive noise for trajectory publishing tasks. \cite{brauer2024time} introduced temporal perturbation strategies to obfuscate trajectories while preserving their analytical utility.

Laplace-based methods have also been adapted for continuous queries in real-time services. \cite{hua2017geo} and \cite{ma2018agent} refined adaptive noise techniques to optimize privacy-utility trade-offs in continuous location reporting. Similarly, \cite{huang2016eppd} developed a proximity testing framework utilizing Laplace noise to ensure privacy during frequent spatial closeness verifications.

Beyond location privacy, federated learning systems have benefited from Laplace-based mechanisms. \cite{galli2023advancing} introduced group-level privacy by adding Laplace noise to aggregated data, and \cite{galli2023group} further refined these techniques to preserve privacy without compromising model performance in collaborative learning scenarios.

The theoretical foundations of the Laplace mechanism have been extended to diverse applications. \cite{koufogiannis2015optimality} and \cite{fernandes2021laplace} demonstrated its utility-optimality for continuous query functions, while \cite{elsalamouny2018optimal} derived optimal noise functions for location privacy. \cite{pankova2022interpreting} analyzed privacy-utility trade-offs for categorical and numerical attributes. \cite{min2024semantic} proposed a semantic Geo-Ind mechanism, adjusting noise levels based on contextual semantics.

For incremental and temporal data, the Laplace mechanism has been adapted to address evolving privacy challenges. \cite{Koufogiannis-JPC2017} introduced incremental data release mechanisms to ensure privacy at each stage of data publication. \cite{xiang2020linear} optimized Laplace noise for temporal queries under metric-based local differential privacy. \cite{kamalaruban2020not} proposed dx-privacy, combining tailored Laplace noise with selective constraints for enhanced privacy protection.

In time-series data, \cite{fan2018time} applied Laplace mechanisms to preserve temporal patterns while ensuring privacy. \cite{brauer2024time} further refined temporal perturbation techniques to improve privacy preservation for trajectory data.

Beyond numerical data, the Laplace mechanism has been applied to text, image, and IoT privacy. For text data, \cite{fernandes2018author} utilized Laplace noise for semantic integrity in author anonymization, while \cite{Fernandes-PST2019} balanced semantic coherence with privacy guarantees. In image privacy, \cite{chen2021perceptual} developed PI-Net, which integrates perceptual distances to obfuscate sensitive visual content. For IoT applications, \cite{liu2018epic} combined Laplace noise with linear programming to mitigate privacy risks in smart home environments.

Crowdsensing and task allocation systems have also leveraged Laplace mechanisms effectively. \cite{huang2019incentivizing} applied Laplace noise to obfuscate user locations in crowdsensing tasks, while \cite{shi2021clap} proposed contract-based mechanisms to ensure fairness in reward distribution. \cite{wang2018personalized} and \cite{To-ICDE2018} developed personalized frameworks for task allocation, optimizing privacy with Laplace-based perturbation techniques.

\section{laplace}
The Laplace mechanism is one of the most widely used data perturbation techniques, originally developed to achieve DP \cite{Dwork-TC2006}. It adds noise $x$ to the output of the query function $f(x)$, i.e., 
$\mathcal{M}(f(x)) = f(x)+x$, where $x$ is sampled from a Laplace distribution with a probability density function
$\mathrm{Lap}(x) = \frac{\epsilon}{2\Delta f} e^{-\frac{\epsilon |x|}{\Delta f}}$. Here, $\Delta f = \sup_{x, y \in \mathcal{X}} |f(x) - f(x')|$ represents the global sensitivity of $f$, which measures the maximum change in the function's output between neighboring datasets.

The Laplace mechanism can naturally adapt to various metrics by appropriately scaling the noise \cite{McSherry-FOCS2007}. Specifically, to extend Laplace for mDP, \cite{Chatzikokolakis-PETS2013} first proposed scaling the noise based on the metric $d_{\mathcal{X}}$ instead of the global sensitivity $\Delta f$, where $d_{\mathcal{X}}(s, x)$ quantifies the distinguishability between $x$ and $x$. The resulting probability density function is expressed as 
$\mathrm{Lap}(x) = \lambda(x) e^{-d_{\mathcal{X}}(s, x)}$.  Following this work, \cite{Andres-CCS2013} 
introducing polar Laplace noise to achieve \emph{geo-indistinguishability (Geo-Ind)}. This approach employs a two-dimensional Laplace distribution centered at the true location $x$, with a probability function given by 
$\Pr\left[\mathcal{M}(x) = y\right] = \frac{\epsilon^2}{2\pi} e^{-\epsilon d(x, y)}$.  

Following \cite{Andres-CCS2013}, many studies have applied the Laplace mechanism for Geo-Ind. \cite{chatzikokolakis2014predictive} tailored Laplace noise for synthetic route generation in smartphone applications. \cite{wang2014entropy} proposed an entropy-minimizing mechanism combining Laplace noise with system feedback control. \cite{arcolezi2021preserving} and \cite{shi2021clap} adapted Geo-Ind for emergency location sharing and incentivized crowdsensing platforms, respectively. \cite{min20213d} extended Geo-Ind to three-dimensional indoor spaces, addressing challenges in high-resolution spatial data. Similarly, \cite{min2023personalized} developed personalized 3D location privacy using distortion-based geo-perturbation techniques. \cite{yang2021blockchain} integrated blockchain with Laplace noise for secure indoor data anonymization.

Beyond basic Geo-Ind, more advanced use cases emerged in trajectory and mobility trace privacy. For mobility traces, \cite{chatzikokolakis2014predictive} generalized Geo-Ind to mobility trace privacy by incorporating predictive correlations across multiple reports. Similarly, \cite{mendes2018effect} examined update frequencies and suggested adaptive noise calibration to balance privacy and utility.

Trajectory data presented a further challenge for privacy preservation, particularly during the COVID-19 pandemic. For example, \cite{li2023accurate} proposed a Laplace-based mechanism for accurate contact tracing. Building on this, \cite{yu2023privacy} developed longitudinal Geo-Ind to protect trajectory privacy in advertising systems. Other studies, such as \cite{li2023privacy} and \cite{duarte2024privacy}, refined these approaches by applying tailored metrics and adaptive noise to trajectory publishing tasks. \cite{brauer2024time} introduced temporal perturbation strategies to obfuscate trajectories while preserving their analytical utility.

Laplace-based approaches also addressed continuous queries in real-time services. For instance, \cite{hua2017geo} and \cite{ma2018agent} refined adaptive noise techniques to optimize privacy-utility trade-offs. Similarly, \cite{huang2016eppd} developed a proximity testing framework that utilized Laplace noise to ensure privacy during frequent spatial closeness verifications.

Expanding beyond location privacy, federated learning systems have also benefited from the Laplace mechanism. In federated learning, \cite{galli2023advancing} proposed Laplace-based mechanisms for group privacy by introducing noise at the group level. \cite{galli2023group} refined these mechanisms to improve collaborative learning privacy without compromising model performance.

The theoretical foundation of the Laplace mechanism has been further developed to reinforce its utility in diverse applications. \cite{koufogiannis2015optimality} and \cite{fernandes2021laplace} demonstrated its utility-optimality for continuous query functions. \cite{elsalamouny2018optimal} derived optimal noise functions for location privacy. \cite{pankova2022interpreting} analyzed privacy-utility trade-offs for categorical and numerical attributes. \cite{min2024semantic} proposed a semantic Geo-Ind mechanism adapting noise levels based on contextual semantics.

In scenarios involving incremental and temporal data, the Laplace mechanism has been adapted to address evolving privacy challenges. \cite{Koufogiannis-JPC2017} introduced incremental data release mechanisms ensuring privacy at each stage. \cite{xiang2020linear} optimized Laplace noise for temporal queries under metric-based local differential privacy. \cite{kamalaruban2020not} developed dx-privacy combining tailored Laplace noise with selective constraints.

For time-series data, \cite{fan2018time} applied Laplace mechanisms to preserve temporal patterns while ensuring privacy. \cite{brauer2024time} further refined temporal perturbation for trajectory data.

Additionally, the Laplace mechanism has been applied across various domains, including text, image, and IoT privacy. For text data, \cite{fernandes2018author} utilized Laplace noise for semantic integrity in author anonymization. \cite{Fernandes-PST2019} balanced semantic coherence with privacy guarantees. \cite{arnold2023guiding} used syntactic structures to guide noise addition.

For image privacy, \cite{chen2021perceptual} developed PI-Net, integrating perceptual distances to obfuscate sensitive content. For IoT, \cite{liu2018epic} combined Laplace noise with linear programming to address privacy risks in smart homes.

Crowdsensing applications and task allocation systems have also incorporated Laplace-based techniques to protect user privacy effectively. \cite{huang2019incentivizing} applied Laplace mechanisms to obfuscate user locations in crowdsensing. \cite{shi2021clap} proposed contract-based mechanisms ensuring fairness in reward distribution. \cite{wang2018personalized} and \cite{To-ICDE2018} developed personalized frameworks optimizing task allocation with Laplace noise.

\section*{Acknowledgments}
We acknowledge the contributions and support of our colleagues and institutions in the preparation of this overview article.

\begin{table*}[h!]
\centering
\footnotesize 
\begin{tabular}{p{5.1cm}|p{4.0cm}|p{6.5cm}}
\hline
References & Perturbation probability distribution $\Pr\left[\mathcal{M}(x) = y\right] $ & Description \\ \hline \hline
PETS 2013 \cite{Chatzikokolakis-PETS2013}& $\lambda e^{-\epsilon d(x, y)}$ 
& Defines a Laplace mechanism over a metric space. The mechanism outputs $x$ with probability proportional to $e^{-\epsilon d(x, y)}$, ensuring metric privacy. \\ \hline

CCS 2013 \cite{Andres-CCS2013}, PETS 2014 \cite{chatzikokolakis2014predictive}, ToDP 2016 \cite{elsalamouny2016differential}, TIFS 2017 \cite{tong2017jointly}, TIFS 2018 \cite{hua2017geo}, CVPR 2021 \cite{chen2021perceptual}, ADMA 2023 \cite{li2023privacy},ToSP 2023 \cite{yu2023balancing}& $\frac{\epsilon^2}{2\pi} e^{-\epsilon \|x - y\|_2}$ 
& Defines the Laplace mechanism for planar geo-indistinguishability, where noise is calibrated to the Euclidean distance $\|x - y\|_2$. \\ \hline

CDC 2014 \cite{wang2014entropy}, BigData 2018 \cite{To-ICDE2018}& $\left(\frac{\epsilon}{2}\right)^n e^{-\epsilon \|x - y\|_1}$
& Extends the Laplace mechanism to $n$-dimensional data. The noise follows a multivariate Laplace distribution calibrated to $\ell_1$-norm sensitivity. \\ \hline

arXiv 2015 \cite{koufogiannis2015optimality}, VLDB 2015 \cite{haney2015design}, JPC 2016 \cite{Koufogiannis-JPC2017}, & $\frac{\epsilon}{2} e^{-\epsilon |x - y|}$ 
& Demonstrates the optimality of Laplace noise under metric privacy constraints. This noise minimizes the mean squared error for $\epsilon$-Lipschitz mechanisms. \\ \hline

POST 2019 \cite{Fernandes-PST2019}& $c_{\epsilon}^n e^{-\epsilon \|x - y\|_2}$, where $c_{\epsilon}^n = \frac{\epsilon^n}{(n-1)!\delta^n}$ 
& Proposes the $n$-dimensional Laplace mechanism with noise added to a vector $x \in \mathbb{R}^n$. The noise is sampled in polar coordinates, where the radius $r$ is drawn from a Gamma distribution $\text{Gam}_\delta^n(r)$ and the direction $p$ is uniformly sampled over the unit hypersphere $B^n$. \\ \hline
ESORICS 2019 \cite{kawamoto2019local}, IET Blockchain 2021 \cite{yang2021blockchain}& $\frac{\epsilon^2}{2\pi} \frac{1}{1 - (1 + \epsilon R)e^{-\epsilon R}} e^{-\epsilon d(x, y)}$ 
& Proposes the Truncated Laplace Mechanism for achieving truncated geo-indistinguishability, confining noise addition to a bounded radius $R$. \\ \hline
ESORICS 2021 \cite{fernandes2021locality}, ICISSP 2023 \cite{galli2023group}& $\frac{1}{C_{\epsilon}} e^{-\epsilon d(x, y)}$, where $C_{\epsilon} = \int_{x \in \mathcal{X}} e^{-\epsilon d(x, y)} dx$ 
& Generalizes the Laplace mechanism to arbitrary metric spaces, achieving $(\epsilon, d_{\mathcal{X}})$-privacy with applications in various domains. \\ \hline
TWC 2022 \cite{min20213d}& $A e^{-\epsilon d_3(x_1, y)}$,  where $d_3(x_1, y)$ is the Euclidean distance in $\mathbb{R}^3$ & Proposes a three-variates Laplacian mechanism centered at $x_1 \in \mathbb{R}^3$. The probability density function (PDF) of the mechanism perturbs $x_1$ to $y$ based on the Euclidean distance $d_3(x_1, y)$. In the spherical coordinate system $(r, \theta, \psi)$, the PDF is reformulated as $D_{\epsilon}(r, \theta, \psi) = \frac{1}{4\pi^2 \epsilon^3} r^2 e^{-\epsilon r}$, where $r$ represents the distance, $\theta$ is the polar angle, and $\psi$ is the azimuth angle. This adaptation simplifies noise generation and improves efficiency for location privacy applications. \\ \hline
CSUS 2024 \cite{brauer2024time} & $\lambda(y) e^{-\epsilon d(x, y)}$, where $\lambda(y) = \frac{\epsilon}{2}$ & Proposes a Laplace mechanism variant for temporal data, where the sensitive information is the timestamp $t$ at which an event occurred. The mechanism perturbs $x$ to $y$ based on the temporal distance $d(x, y) = |x - y|$. This $\epsilon$-temporal trajectory indistinguishability ensures privacy by calibrating noise to the temporal domain, with a scaling function $\lambda(y)$. \\ \hline
\end{tabular}
\caption{Summary of Laplace Mechanism \qiu{add references in the table}}
\label{tab:privacy_summary}
\end{table*}

\begin{table*}[h!]
\centering
\footnotesize 

\begin{tabular}{p{3.5cm}|p{4.0cm}|p{9.5cm}}
\hline
References & Perturbation probability distribution $\Pr\left[\mathcal{M}(x) = y\right]$ & Description \\ \hline \hline
PETS 2015 \cite{chatzikokolakis2015constructing}, NDSS 2017 \cite{Yu-NDSS2017}, IJCAI 2019 \cite{fioretto2019privacy}, ToVT 2019 \cite{dong2019preserving}, TMC 2022 \cite{niu2020eclipse}& $c_x e^{-\frac{1}{2} d(x, y)}$, where $c_x = \left(\sum_{y' \in \mathcal{Y}} e^{-\frac{1}{2} d_{\mathcal{X}}(x, y')}\right)^{-1}$ & Proposes a variant of the Exponential Mechanism for arbitrary distinguishability metrics $d_{\mathcal{X}}$. The mechanism outputs $z$ with probability proportional to $e^{-\frac{1}{2} d(x, y)}$, satisfying $d_{\mathcal{X}}$-privacy. This is the earliest formulation of the Exponential Mechanism in a normalized probability form and serves as the foundation for subsequent adaptations. \\ \hline

TDSC 2019 \cite{gursoy2019secure}& $\frac{e^{-\frac{\alpha  d(x, y)}{2}}}{\sum_{y' \in \mathcal{Y}} e^{-\frac{\alpha  d(x, y')}{2}}}$ & Proposes the Exponential Mechanism for achieving $\alpha$-Condensed Local Differential Privacy (CLDP). Unlike traditional Exponential Mechanisms, this method adjusts the sensitivity to fit local privacy needs within a metric space. The normalization factor ensures probabilities sum to 1 while introducing new privacy concepts specifically tailored for local privacy. \\ 
\hline
UAI 2022 \cite{imola2022balancing}& $\frac{Y_y  e^{-\frac{\epsilon  d(x, y)}{2}}}{\sum_{y' \in \mathcal{Y}} Y_{y'}  e^{-\frac{\epsilon  d(x, y')}{2}}}$ & Proposes a Weighted Exponential Mechanism (ExpMech\(_Y\)) to balance the probabilities assigned to outputs by introducing positive weights $Y = (Y_1, Y_2, \dots, Y_n) \in (\mathbb{R}^+)^n$. This mechanism solves the ``black hole'' problem in dense metric spaces by assigning weights to outputs, achieving better utility while maintaining $\epsilon$-privacy. \\ \hline
SDM 2023 \cite{carvalho2023tem}& $\propto  \exp\left(-\frac{\epsilon}{2}  d(x, y)\right)$, if $z \in L_x$; \newline $\propto \exp\left(-\gamma + \frac{2 \ln(|\mathcal{Y}| - |L_x|)}{\epsilon}\right)$, if  $z = \perp$. & Proposes the Metric Truncated Exponential Mechanism (TEM), which adapts the Exponential Mechanism for metric-DP with truncation and Gumbel noise. The novelty lies in truncating the candidate output space to a subset $L_x = \{z \in \mathcal{Y} \mid d(x, y) \leq \gamma\}$ and handling out-of-threshold elements through a special $\perp$ element. By adding Gumbel noise, this mechanism improves utility for text-based metric-DP applications. \\ \hline

\end{tabular}
\caption{Summary of Exponential Mechanism}
\label{tab:privacy_summary_1}
\end{table*}

\begin{table*}
\scriptsize    
\centering
\begin{tabular}{|p{4.1cm}|p{3cm}|p{5cm}|p{4cm}|}
\hline
\textbf{Paper} & \textbf{Distance} & \textbf{Formula} & \textbf{Explanation} \\ \hline
PETS 2013 \cite{Chatzikokolakis-PETS2013}, PETS 2014 \cite{chatzikokolakis2014predictive}, CCS 2014 \cite{fawaz2014location} & Euclidean Distance & $d(x, y) = \sqrt{\sum_{i=1}^n (x_i - y_i)^2}$ & Measures the straight-line distance between two points in $n$-dimensional space. Commonly used in geo-location privacy to quantify spatial proximity. \\ \hline

CDC 2014 \cite{wang2014entropy}& Manhattan Distance & $d(x, y) = \sum_{i=1}^n |x_i - y_i|$ & Measures the distance between two points in a grid-based layout, summing absolute differences along each axis. Used in privacy scenarios where the grid structure is natural. \\ \hline

VLDB 2015 \cite{haney2015design}& Graph Distance & $d_G(x, y)$ & Defines the distance between two nodes $x$ and $y$ in a graph $G$ as the length of the shortest path connecting them. This graph is a policy graph that encodes domain-specific policies about node relationships. \\ \hline

INFOCOM 2017 \cite{wang2017local}& Categorical Distance & $d(x, y) = |x - y|$ & For ordinal categories, the distance is determined by the absolute difference between the indices of two categories. This assumes a linear ordering of the categories. \\ \hline

arXiv 2018 \cite{fernandes2018author}, CSF 2018\cite{alvim2018metric}, POST 2019 \cite{Fernandes-PST2019}& Word Mover's Distance (Kantorovich distance) & $d_w(x, y) = \min_{T \geq 0} \sum_{i,j} T_{ij} C_{ij}$, subject to $\sum_j T_{ij} = \frac{1}{a} \, \forall i$, $\sum_i T_{ij} = \frac{1}{b} \, \forall j$ & Measures the cost of transforming one distribution $x$ into another $y$. $T_{ij}$ denotes how much of mass $i$ in $x$ is moved to mass $j$ in $y$, and $C_{ij}$ represents the cost of moving mass $i$ to mass $j$. \\ \hline

FOCS 2018 \cite{borgs2018revealing} & Node-Distance & $d(x, y) = \min \{ \text{number of nodes in } x \text{ to modify to obtain } y \}$ & Measures the minimum number of nodes in graph $x$ that need to be modified to transform it into graph $y$. \\ \hline

DBSec 2019 \cite{takagi2019geo}, TMC 2020\cite{qiu2020location}, ESORICS 2020 \cite{cao2020pglp}, TITS 2022 \cite{ma2022personalized}, SIGSPATIAL 2022 \cite{qiu2022trafficadaptor} & Shortest Path Distance & $d_s(x, y) = \min \{\text{length of all paths connecting } x \text{ and } y \}$ & Represents the shortest path length between nodes $x$ and $y$ in a weighted, undirected graph $(V, E)$. This distance is based on the road network and considers the minimum cumulative weight of edges connecting the two nodes. \\ \hline

ICDM 2019 \cite{feyisetan2019leveraging} & Hyperbolic Distance & $d(x, y) = \operatorname{arcosh}\left(1 + 2\frac{\lVert x - y \rVert^2}{(1 - \lVert x \rVert^2)(1 - \lVert y \rVert^2)}\right)$ & Defined in the Poincaré ball model of hyperbolic space, where $\lVert \cdot \rVert$ represents the Euclidean norm. \\ \hline

ESORICS 2019 \cite{kawamoto2019local}& Wasserstein Distance & $W_\infty(x, y) = \min_{\gamma \in \text{cp}(x, y)} \max_{(x', y') \in \text{supp}(\gamma)} d(x', y')$ & Measures the highest cost of transporting probability mass between two distributions $x$ and $y$. \\ \hline

arXiv 2020 \cite{xu2020differentially}, TKDE 2023 \cite{zhao2022geo} & Regularized Mahalanobis Norm & $\|x\|_{RM} = \sqrt{x^\top \{ \lambda \Sigma + (1 - \lambda) \mathbf{I}_m \}^{-1} x}$ & Defines a regularized Mahalanobis distance metric for elliptical noise calibration. \\ \hline

ICME 2020 \cite{han2020voice}, CCS 2021 \cite{weggenmann2021differential}, ESORICS 2021 \cite{fernandes2021locality}& Angular Distance & $d(x, y) = \frac{\arccos(\cos \text{similarity } \langle x, y \rangle)}{\pi}$ & Defines the angular distance between two vectors $x$ and $y$. \\ \hline

CVPR 2021 \cite{chen2021perceptual} & Perceptual Distance & $d(x, y) = \|x - y\|_2$ & Defines the perceptual distance as the Euclidean distance between two latent codes $x$ and $y$ in the latent space learned by a GAN model, capturing perceptual similarity between images. \\ \hline

SEBD 2023 \cite{boninsegna2023locality} & Fréchet Distance & $d(x, y) = \min_{T \in \mathcal{T}} \max_{(i, j) \in T} \|x_i - y_j\|$ & Defines the Fréchet distance between two curves $x$ and $y$. \\ \hline

arXiv 2024 \cite{brauer2024time} & Temporal Distance & $d(x, y) = |x - y|$ & Defines the temporal distance between two timestamps $x$ and $y$. \\ \hline
\end{tabular}
\caption{Summary of Metric Definitions in Metric Differential Privacy}
\label{tab:metric-dp-distances}
\end{table*}
}

\newpage 


\begin{thebibliography}{100}
\providecommand{\url}[1]{#1}
\csname url@samestyle\endcsname
\providecommand{\newblock}{\relax}
\providecommand{\bibinfo}[2]{#2}
\providecommand{\BIBentrySTDinterwordspacing}{\spaceskip=0pt\relax}
\providecommand{\BIBentryALTinterwordstretchfactor}{4}
\providecommand{\BIBentryALTinterwordspacing}{\spaceskip=\fontdimen2\font plus
\BIBentryALTinterwordstretchfactor\fontdimen3\font minus \fontdimen4\font\relax}
\providecommand{\BIBforeignlanguage}[2]{{%
\expandafter\ifx\csname l@#1\endcsname\relax
\typeout{** WARNING: IEEEtran.bst: No hyphenation pattern has been}%
\typeout{** loaded for the language `#1'. Using the pattern for}%
\typeout{** the default language instead.}%
\else
\language=\csname l@#1\endcsname
\fi
#2}}
\providecommand{\BIBdecl}{\relax}
\BIBdecl

\bibitem{dwork2014algorithmic}
C.~Dwork, A.~Roth \emph{et~al.}, ``The algorithmic foundations of differential privacy,'' \emph{Foundations and Trends{\textregistered} in Theoretical Computer Science}, vol.~9, no. 3--4, pp. 211--407, 2014.

\bibitem{imola2022balancing}
J.~Imola, S.~Kasiviswanathan, S.~White, A.~Aggarwal, and N.~Teissier, ``Balancing utility and scalability in metric differential privacy,'' in \emph{Uncertainty in Artificial Intelligence}.\hskip 1em plus 0.5em minus 0.4em\relax PMLR, 2022, pp. 885--894.

\bibitem{Chatzikokolakis-PETS2013}
K.~Chatzikokolakis, M.~E. Andr{\'e}s, N.~E. Bordenabe, and C.~Palamidessi, ``Broadening the scope of differential privacy using metrics,'' in \emph{international symposium on privacy enhancing technologies symposium}.\hskip 1em plus 0.5em minus 0.4em\relax Springer, 2013, pp. 82--102.

\bibitem{Andres-CCS2013}
M.~E. Andr{\'e}s, N.~E. Bordenabe, K.~Chatzikokolakis, and C.~Palamidessi, ``Geo-indistinguishability: Differential privacy for location-based systems,'' in \emph{Proceedings of the 2013 ACM SIGSAC conference on Computer \& communications security}, 2013, pp. 901--914.

\bibitem{feyisetan2021private}
O.~Feyisetan and S.~Kasiviswanathan, ``Private release of text embedding vectors,'' in \emph{Proceedings of the First Workshop on Trustworthy Natural Language Processing}, 2021, pp. 15--27.

\bibitem{han2020voice}
Y.~Han, S.~Li, Y.~Cao, Q.~Ma, and M.~Yoshikawa, ``Voice-indistinguishability: Protecting voiceprint in privacy-preserving speech data release,'' in \emph{2020 IEEE International Conference on Multimedia and Expo (ICME)}.\hskip 1em plus 0.5em minus 0.4em\relax IEEE, 2020, pp. 1--6.

\bibitem{fan2019practical}
L.~Fan, ``Practical image obfuscation with provable privacy,'' in \emph{2019 IEEE international conference on multimedia and expo (ICME)}.\hskip 1em plus 0.5em minus 0.4em\relax IEEE, 2019, pp. 784--789.

\bibitem{chen2021perceptual}
J.-W. Chen, L.-J. Chen, C.-M. Yu, and C.-S. Lu, ``Perceptual indistinguishability-net (pi-net): Facial image obfuscation with manipulable semantics,'' in \emph{Proceedings of the IEEE/CVF Conference on Computer Vision and Pattern Recognition}, 2021, pp. 6478--6487.

\bibitem{zhao2022survey}
Y.~Zhao and J.~Chen, ``A survey on differential privacy for unstructured data content,'' \emph{ACM Computing Surveys (CSUR)}, vol.~54, no. 10s, pp. 1--28, 2022.

\bibitem{zhao2024scenario}
Y.~Zhao, J.~T. Du, and J.~Chen, ``Scenario-based adaptations of differential privacy: A technical survey,'' \emph{ACM Computing Surveys}, vol.~56, no.~8, pp. 1--39, 2024.

\bibitem{Dwork-TC2006}
C.~Dwork, F.~McSherry, K.~Nissim, and A.~Smith, ``Calibrating noise to sensitivity in private data analysis,'' in \emph{Theory of Cryptography: Third Theory of Cryptography Conference, TCC 2006, New York, NY, USA, March 4-7, 2006. Proceedings 3}.\hskip 1em plus 0.5em minus 0.4em\relax Springer, 2006, pp. 265--284.

\bibitem{koufogiannis2015optimality}
F.~Koufogiannis, S.~Han, and G.~J. Pappas, ``Optimality of the laplace mechanism in differential privacy,'' \emph{arXiv preprint arXiv:1504.00065}, 2015.

\bibitem{feyisetan2020privacy}
O.~Feyisetan, B.~Balle, T.~Drake, and T.~Diethe, ``Privacy-and utility-preserving textual analysis via calibrated multivariate perturbations,'' in \emph{Proceedings of the 13th international conference on web search and data mining}, 2020, pp. 178--186.

\bibitem{dharangutte2023integer}
P.~Dharangutte, J.~Gao, R.~Gong, and F.-Y. Yu, ``Integer subspace differential privacy,'' in \emph{Proceedings of the AAAI Conference on Artificial Intelligence}, vol.~37, no.~6, 2023, pp. 7349--7357.

\bibitem{KasiviswanathanJoC}
\BIBentryALTinterwordspacing
S.~P. Kasiviswanathan, H.~K. Lee, K.~Nissim, S.~Raskhodnikova, and A.~D. Smith, ``What can we learn privately?'' \emph{CoRR}, vol. abs/0803.0924, 2008. [Online]. Available: \url{http://arxiv.org/abs/0803.0924}
\BIBentrySTDinterwordspacing

\bibitem{chatzikokolakis2014predictive}
K.~Chatzikokolakis, C.~Palamidessi, and M.~Stronati, ``A predictive differentially-private mechanism for mobility traces,'' in \emph{Privacy Enhancing Technologies: 14th International Symposium, PETS 2014, Amsterdam, The Netherlands, July 16-18, 2014. Proceedings 14}.\hskip 1em plus 0.5em minus 0.4em\relax Springer, 2014, pp. 21--41.

\bibitem{fawaz2014location}
K.~Fawaz and K.~G. Shin, ``Location privacy protection for smartphone users,'' in \emph{Proceedings of the 2014 ACM SIGSAC Conference on Computer and Communications Security}, 2014, pp. 239--250.

\bibitem{feyisetan2019leveraging}
O.~Feyisetan, T.~Diethe, and T.~Drake, ``Leveraging hierarchical representations for preserving privacy and utility in text,'' in \emph{2019 IEEE International Conference on Data Mining (ICDM)}.\hskip 1em plus 0.5em minus 0.4em\relax IEEE, 2019, pp. 210--219.

\bibitem{wang2014entropy}
Y.~Wang, Z.~Huang, S.~Mitra, and G.~E. Dullerud, ``Entropy-minimizing mechanism for differential privacy of discrete-time linear feedback systems,'' in \emph{53rd IEEE conference on decision and control}.\hskip 1em plus 0.5em minus 0.4em\relax IEEE, 2014, pp. 2130--2135.

\bibitem{fernandes2018author}
N.~Fernandes, M.~Dras, and A.~McIver, ``Author obfuscation using generalised differential privacy,'' \emph{arXiv preprint arXiv:1805.08866}, 2018.

\bibitem{Fernandes-PST2019}
------, ``Generalised differential privacy for text document processing,'' in \emph{Principles of Security and Trust: 8th International Conference, POST 2019, Held as Part of the European Joint Conferences on Theory and Practice of Software, ETAPS 2019, Prague, Czech Republic, April 6--11, 2019, Proceedings 8}.\hskip 1em plus 0.5em minus 0.4em\relax Springer International Publishing, 2019, pp. 123--148.

\bibitem{borgs2018revealing}
C.~Borgs, J.~Chayes, A.~Smith, and I.~Zadik, ``Revealing network structure, confidentially: Improved rates for node-private graphon estimation,'' in \emph{2018 IEEE 59th Annual Symposium on Foundations of Computer Science (FOCS)}.\hskip 1em plus 0.5em minus 0.4em\relax IEEE, 2018, pp. 533--543.

\bibitem{haney2014design}
S.~Haney, A.~Machanavajjhala, and B.~Ding, ``Design of policy-aware differentially private algorithms,'' \emph{arXiv preprint arXiv:1404.3722}, 2014.

\bibitem{qiu2020location}
C.~Qiu, A.~Squicciarini, C.~Pang, N.~Wang, and B.~Wu, ``Location privacy protection in vehicle-based spatial crowdsourcing via geo-indistinguishability,'' \emph{IEEE Transactions on Mobile Computing}, vol.~21, no.~7, pp. 2436--2450, 2020.

\bibitem{cao2020pglp}
Y.~Cao, Y.~Xiao, S.~Takagi, L.~Xiong, M.~Yoshikawa, Y.~Shen, J.~Liu, H.~Jin, and X.~Xu, ``Pglp: Customizable and rigorous location privacy through policy graph,'' in \emph{European Symposium on Research in Computer Security}.\hskip 1em plus 0.5em minus 0.4em\relax Springer, 2020, pp. 655--676.

\bibitem{ma2022personalized}
B.~Ma, X.~Wang, W.~Ni, and R.~P. Liu, ``Personalized location privacy with road network-indistinguishability,'' \emph{IEEE Transactions on Intelligent Transportation Systems}, vol.~23, no.~11, pp. 20\,860--20\,872, 2022.

\bibitem{kawamoto2019local}
Y.~Kawamoto and T.~Murakami, ``Local obfuscation mechanisms for hiding probability distributions,'' in \emph{European Symposium on Research in Computer Security}.\hskip 1em plus 0.5em minus 0.4em\relax Springer, 2019, pp. 128--148.

\bibitem{xu2020differentially}
Z.~Xu, A.~Aggarwal, O.~Feyisetan, and N.~Teissier, ``A differentially private text perturbation method using a regularized mahalanobis metric,'' \emph{arXiv preprint arXiv:2010.11947}, 2020.

\bibitem{zhao2022geo}
Y.~Zhao, D.~Yuan, J.~T. Du, and J.~Chen, ``Geo-ellipse-indistinguishability: Community-aware location privacy protection for directional distribution,'' \emph{IEEE Transactions on Knowledge and Data Engineering}, vol.~35, no.~7, pp. 6957--6967, 2022.

\bibitem{weggenmann2021differential}
B.~Weggenmann and F.~Kerschbaum, ``Differential privacy for directional data,'' in \emph{Proceedings of the 2021 ACM SIGSAC Conference on Computer and Communications Security}, 2021, pp. 1205--1222.

\bibitem{fernandes2021locality}
N.~Fernandes, Y.~Kawamoto, and T.~Murakami, ``Locality sensitive hashing with extended differential privacy,'' in \emph{European Symposium on Research in Computer Security}.\hskip 1em plus 0.5em minus 0.4em\relax Springer, 2021, pp. 563--583.

\bibitem{boninsegna2023locality}
F.~Boninsegna \emph{et~al.}, ``Locality sensitive hashing of trajectories under local differential privacy.'' in \emph{SEBD}, 2023, pp. 681--687.

\bibitem{brauer2024time}
A.~Brauer, V.~M{\"a}kinen, L.~Ruotsalainen, and J.~Oksanen, ``Time will not tell: Temporal approaches for privacy-preserving trajectory publishing,'' \emph{Computers, Environment and Urban Systems}, vol. 112, p. 102154, 2024.

\bibitem{yang2024local}
M.~Yang, T.~Guo, T.~Zhu, I.~Tjuawinata, J.~Zhao, and K.-Y. Lam, ``Local differential privacy and its applications: A comprehensive survey,'' \emph{Computer Standards \& Interfaces}, vol.~89, p. 103827, 2024.

\bibitem{McSherry-FOCS2007}
F.~McSherry and K.~Talwar, ``Mechanism design via differential privacy,'' in \emph{48th Annual IEEE Symposium on Foundations of Computer Science (FOCS'07)}.\hskip 1em plus 0.5em minus 0.4em\relax IEEE, 2007, pp. 94--103.

\bibitem{hua2017geo}
J.~Hua~et al., ``A geo-indistinguishable location perturbation mechanism for location-based services supporting frequent queries,'' \emph{TIFS}, 2017.

\bibitem{ma2018agent}
X.~Ma~et al., ``Agent: An adaptive geo-indistinguishable mechanism for continuous location-based service,'' \emph{Peer-to-Peer Netw. Appl.}, 2018.

\bibitem{Chen-IoTJ2022}
R.~Chen, L.~Li, Y.~Ma, Y.~Gong, Y.~Guo, T.~Ohtsuki, and M.~Pan, ``Constructing mobile crowdsourced covid-19 vulnerability map with geo-indistinguishability,'' \emph{IEEE Internet of Things Journal}, vol.~9, no.~18, pp. 17\,403--17\,416, 2022.

\bibitem{mendes2018effect}
R.~Mendes and J.~Vilela, ``On the effect of update frequency on geo-indistinguishability of mobility traces,'' in \emph{Proceedings of the 11th ACM Conference on Security \& Privacy in Wireless and Mobile Networks}, 2018, pp. 271--276.

\bibitem{to2018privacy}
H.~To, C.~Shahabi, and L.~Xiong, ``Privacy-preserving online task assignment in spatial crowdsourcing with untrusted server,'' in \emph{2018 IEEE 34th international conference on data engineering (ICDE)}.\hskip 1em plus 0.5em minus 0.4em\relax IEEE, 2018, pp. 833--844.

\bibitem{yang2021blockchain}
C.~Yang, X.~Ju, E.~Liu, Y.~Geng, and R.~Wang, ``Blockchain-based indoor location paging and answering service with truncated-geo-indistinguishability,'' \emph{IET Blockchain}, vol.~1, no. 2-4, pp. 105--117, 2021.

\bibitem{huang2019incentivizing}
P.~Huang, X.~Zhang, L.~Guo, and M.~Li, ``Incentivizing crowdsensing-based noise monitoring with differentially-private locations,'' \emph{IEEE Transactions on Mobile Computing}, vol.~20, no.~2, pp. 519--532, 2019.

\bibitem{li2023privacy}
F.~Li, J.~Dong, M.~Chen, and P.~Li, ``A privacy preserving method for trajectory data publishing based on geo-indistinguishability,'' in \emph{International Conference on Advanced Data Mining and Applications}.\hskip 1em plus 0.5em minus 0.4em\relax Springer, 2023, pp. 633--647.

\bibitem{li2023accurate}
M.~Li, Y.~Zeng, L.~Zheng, L.~Chen, and Q.~Li, ``Accurate and efficient trajectory-based contact tracing with secure computation and geo-indistinguishability,'' in \emph{International Conference on Database Systems for Advanced Applications}.\hskip 1em plus 0.5em minus 0.4em\relax Springer, 2023, pp. 300--316.

\bibitem{duarte2024privacy}
G.~Duarte~et al., ``A privacy-aware remapping mechanism for location data,'' in \emph{SAC}, 2024.

\bibitem{min2023personalized}
M.~Min, H.~Zhu, J.~Ding, S.~Li, L.~Xiao, M.~Pan, and Z.~Han, ``Personalized 3d location privacy protection with differential and distortion geo-perturbation,'' \emph{IEEE Transactions on Dependable and Secure Computing}, vol.~21, no.~4, pp. 3629--3643, 2023.

\bibitem{yu2023privacy}
L.~Yu, S.~Zhang, Y.~Meng, S.~Du, Y.~Chen, Y.~Ren, and H.~Zhu, ``Privacy-preserving location-based advertising via longitudinal geo-indistinguishability,'' \emph{IEEE Transactions on Mobile Computing}, vol.~23, no.~8, pp. 8256--8273, 2023.

\bibitem{al2018adaptive}
R.~Al-Dhubhani and J.~M. Cazalas, ``An adaptive geo-indistinguishability mechanism for continuous lbs queries,'' \emph{Wireless Networks}, vol.~24, no.~8, pp. 3221--3239, 2018.

\bibitem{cunha2019clustering}
M.~Cunha, R.~Mendes, and J.~P. Vilela, ``Clustering geo-indistinguishability for privacy of continuous location traces,'' in \emph{2019 4th International Conference on Computing, Communications and Security (ICCCS)}.\hskip 1em plus 0.5em minus 0.4em\relax IEEE, 2019, pp. 1--8.

\bibitem{fernandes2021laplace}
N.~Fernandes, A.~McIver, and C.~Morgan, ``The laplace mechanism has optimal utility for differential privacy over continuous queries,'' in \emph{2021 36th Annual ACM/IEEE Symposium on Logic in Computer Science (LICS)}.\hskip 1em plus 0.5em minus 0.4em\relax IEEE, 2021, pp. 1--12.

\bibitem{oyaGeoIndLooking}
\BIBentryALTinterwordspacing
S.~Oya, C.~Troncoso, and F.~P\'{e}rez-Gonz\'{a}lez, ``Is geo-indistinguishability what you are looking for?'' in \emph{Proc. of the 2017 on Workshop on Privacy in the Electronic Society}, ser. WPES '17.\hskip 1em plus 0.5em minus 0.4em\relax New York, NY, USA: Association for Computing Machinery, 2017, p. 137–140. [Online]. Available: \url{https://doi.org/10.1145/3139550.3139555}
\BIBentrySTDinterwordspacing

\bibitem{chatzikokolakis2017efficient}
K.~Chatzikokolakis, E.~Elsalamouny, and C.~Palamidessi, ``Efficient utility improvement for location privacy,'' \emph{Proceedings on Privacy Enhancing Technologies}, 2017.

\bibitem{pankova2022interpreting}
A.~Pankova and P.~Laud, ``Interpreting epsilon of differential privacy in terms of advantage in guessing or approximating sensitive attributes,'' in \emph{2022 IEEE 35th Computer Security Foundations Symposium (CSF)}.\hskip 1em plus 0.5em minus 0.4em\relax IEEE, 2022, pp. 96--111.

\bibitem{Koufogiannis-JPC2017}
F.~Koufogiannis, S.~Han, and G.~J. Pappas, ``Gradual release of sensitive data under differential privacy,'' \emph{Journal of Privacy and Confidentiality}, vol.~7, no.~2, 2016.

\bibitem{koufogiannis2016location}
F.~Koufogiannis and G.~J. Pappas, ``Location-dependent privacy,'' in \emph{2016 IEEE 55th Conference on Decision and Control (CDC)}.\hskip 1em plus 0.5em minus 0.4em\relax IEEE, 2016, pp. 7586--7591.

\bibitem{AhujaEDBT2019}
R.~Ahuja, G.~Ghinita, and C.~Shahabi, ``A utility-preserving and scalable technique for protecting location data with geo-indistinguishability,'' in \emph{22nd International Conference on Extending Database Technology, EDBT 2019}.\hskip 1em plus 0.5em minus 0.4em\relax OpenProceedings. org, 2019, pp. 217--228.

\bibitem{min2024semantic}
M.~Min, H.~Zhu, S.~Li, H.~Zhang, L.~Xiao, M.~Pan, and Z.~Han, ``Semantic adaptive geo-indistinguishability for location privacy protection in mobile networks,'' \emph{IEEE Transactions on Vehicular Technology}, 2024.

\bibitem{min20213d}
M.~Min, L.~Xiao, J.~Ding, H.~Zhang, S.~Li, M.~Pan, and Z.~Han, ``3d geo-indistinguishability for indoor location-based services,'' \emph{IEEE Transactions on Wireless Communications}, vol.~21, no.~7, pp. 4682--4694, 2021.

\bibitem{fathalizadeh2023indoor}
A.~Fathalizadeh, V.~Moghtadaiee, and M.~Alishahi, ``Indoor geo-indistinguishability: Adopting differential privacy for indoor location data protection,'' \emph{IEEE Transactions on Emerging Topics in Computing}, vol.~12, no.~1, pp. 293--306, 2023.

\bibitem{eltarjaman2017location}
W.~Eltarjaman, R.~Dewri, and R.~Thurimella, ``Location privacy for rank-based geo-query systems,'' \emph{Proceedings on Privacy Enhancing Technologies}, 2017.

\bibitem{kamalaruban2020not}
P.~Kamalaruban, V.~Perrier, H.~J. Asghar, and M.~A. Kaafar, ``Not all attributes are created equal: dx-private mechanisms for linear queries,'' \emph{Proceedings on Privacy Enhancing Technologies}, 2020.

\bibitem{arnold2023guiding}
S.~Arnold, D.~Yesilbas, and S.~Weinzierl, ``Guiding text-to-text privatization by syntax,'' in \emph{Proceedings of the 3rd Workshop on Trustworthy Natural Language Processing (TrustNLP 2023)}, 2023, pp. 151--162.

\bibitem{fioretto2019privacy}
F.~Fioretto, T.~W. Mak, and P.~Van~Hentenryck, ``Privacy-preserving obfuscation of critical infrastructure networks,'' \emph{arXiv preprint arXiv:1905.09778}, 2019.

\bibitem{fan2018time}
L.~Fan and L.~Bonomi, ``Time series sanitization with metric-based privacy,'' in \emph{2018 IEEE International Congress on Big Data (BigData Congress)}.\hskip 1em plus 0.5em minus 0.4em\relax IEEE, 2018, pp. 264--267.

\bibitem{tong2017jointly}
W.~Tong, J.~Hua, and S.~Zhong, ``A jointly differentially private scheduling protocol for ridesharing services,'' \emph{IEEE Transactions on Information Forensics and Security}, vol.~12, no.~10, pp. 2444--2456, 2017.

\bibitem{huang2016eppd}
C.~Huang, R.~Lu, H.~Zhu, J.~Shao, A.~Alamer, and X.~Lin, ``Eppd: Efficient and privacy-preserving proximity testing with differential privacy techniques,'' in \emph{2016 IEEE International Conference on Communications (ICC)}.\hskip 1em plus 0.5em minus 0.4em\relax IEEE, 2016, pp. 1--6.

\bibitem{shi2021clap}
X.~Shi, D.~Yu, M.~Fu, and W.-A. Zhang, ``Clap: A contract-based incentive mechanism for cooperative localization balancing localization accuracy and location privacy,'' \emph{IEEE Internet of Things Journal}, vol.~9, no.~9, pp. 6678--6687, 2021.

\bibitem{galli2023group}
F.~Galli, S.~Biswas, K.~Jung, T.~Cucinotta, C.~Palamidessi \emph{et~al.}, ``Group privacy for personalized federated learning,'' \emph{ICISSP}, vol.~1, pp. 252--263, 2023.

\bibitem{galli2023advancing}
F.~Galli~et al., ``Advancing personalized federated learning: Group privacy, fairness, and beyond,'' \emph{SN Computer Science}, 2023.

\bibitem{chatzikokolakis2015constructing}
K.~Chatzikokolakis, C.~Palamidessi, and M.~Stronati, ``Constructing elastic distinguishability metrics for location privacy,'' \emph{Proceedings on Privacy Enhancing Technologies}, 2015.

\bibitem{takagi2023geo}
S.~Takagi, Y.~Cao, Y.~Asano, and M.~Yoshikawa, ``Geo-graph-indistinguishability: Location privacy on road networks with differential privacy,'' \emph{IEICE TRANSACTIONS on Information and Systems}, vol. 106, no.~5, pp. 877--894, 2023.

\bibitem{ma2022new}
B.~Ma, X.~Lin, X.~Wang, B.~Liu, Y.~He, W.~Ni, and R.~P. Liu, ``New cloaking region obfuscation for road network-indistinguishability and location privacy,'' in \emph{Proceedings of the 25th International Symposium on Research in Attacks, Intrusions and Defenses}, 2022, pp. 160--170.

\bibitem{Yu-NDSS2017}
L.~Yu, L.~Liu, and C.~Pu, ``Dynamic differential location privacy with personalized error bounds.'' in \emph{NDSS}, vol.~17, 2017, pp. 1--15.

\bibitem{shun2021differential}
Z.~Shun, D.~Benfei, C.~Zhili, and Z.~Hong, ``On the differential privacy of dynamic location obfuscation with personalized error bounds,'' \emph{arXiv preprint arXiv:2101.12602}, 2021.

\bibitem{zhang2024dpive}
S.~Zhang, P.~Lan, B.~Duan, Z.~Chen, H.~Zhong, and N.~N. Xiong, ``Dpive: a regionalized location obfuscation scheme with personalized privacy levels,'' \emph{ACM Transactions on Sensor Networks}, vol.~20, no.~2, pp. 1--26, 2024.

\bibitem{gursoy2019secure}
M.~E. Gursoy, A.~Tamersoy, S.~Truex, W.~Wei, and L.~Liu, ``Secure and utility-aware data collection with condensed local differential privacy,'' \emph{IEEE Transactions on Dependable and Secure Computing}, vol.~18, no.~5, pp. 2365--2378, 2019.

\bibitem{ren2022distpreserv}
Y.~Ren, X.~Li, Y.~Miao, R.~H. Deng, J.~Weng, S.~Ma, and J.~Ma, ``Distpreserv: Maintaining user distribution for privacy-preserving location-based services,'' \emph{IEEE Transactions on Mobile Computing}, vol.~22, no.~6, pp. 3287--3302, 2022.

\bibitem{dong2019preserving}
X.~Dong~et al., ``Preserving geo-indistinguishability of the primary user in dynamic spectrum sharing,'' \emph{IEEE Trans. on Vehicular Technology}, 2019.

\bibitem{niu2020eclipse}
B.~Niu, Y.~Chen, Z.~Wang, F.~Li, B.~Wang, and H.~Li, ``Eclipse: Preserving differential location privacy against long-term observation attacks,'' \emph{IEEE Transactions on Mobile Computing}, vol.~21, no.~1, pp. 125--138, 2020.

\bibitem{li2020perturbation}
X.~Li, Y.~Ren, L.~T. Yang, N.~Zhang, B.~Luo, J.~Weng, and X.~Liu, ``Perturbation-hidden: Enhancement of vehicular privacy for location-based services in internet of vehicles,'' \emph{IEEE Transactions on Network Science and Engineering}, vol.~8, no.~3, pp. 2073--2086, 2020.

\bibitem{wang2017local}
S.~Wang, Y.~Nie, P.~Wang, H.~Xu, W.~Yang, and L.~Huang, ``Local private ordinal data distribution estimation,'' in \emph{IEEE INFOCOM 2017-IEEE Conference on Computer Communications}.\hskip 1em plus 0.5em minus 0.4em\relax IEEE, 2017, pp. 1--9.

\bibitem{yue2021differential}
X.~Yue, M.~Du, T.~Wang, Y.~Li, H.~Sun, and S.~S. Chow, ``Differential privacy for text analytics via natural text sanitization,'' in \emph{Findings of the Association for Computational Linguistics: ACL-IJCNLP 2021}, 2021, pp. 3853--3866.

\bibitem{feyisetan2020research}
O.~Feyisetan, A.~Aggarwal, Z.~Xu, and N.~Teissier, ``Research challenges in designing differentially private text generation mechanisms,'' \emph{arXiv preprint arXiv:2012.05403}, 2020.

\bibitem{raskhodnikova2016lipschitz}
S.~Raskhodnikova and A.~Smith, ``Lipschitz extensions for node-private graph statistics and the generalized exponential mechanism,'' in \emph{2016 IEEE 57th Annual Symposium on Foundations of Computer Science (FOCS)}.\hskip 1em plus 0.5em minus 0.4em\relax IEEE, 2016, pp. 495--504.

\bibitem{roy2022strengthening}
A.~Roy~Chowdhury, B.~Ding, S.~Jha, W.~Liu, and J.~Zhou, ``Strengthening order preserving encryption with differential privacy,'' in \emph{Proceedings of the 2022 ACM SIGSAC Conference on Computer and Communications Security}, 2022, pp. 2519--2533.

\bibitem{zhang2022geo}
S.~Zhang, T.~Zhang, Z.~Chen, and N.~Xiong, ``Geo-moea: A multi-objective evolutionary algorithm with geo-obfuscation for mobile crowdsourcing workers,'' \emph{arXiv preprint arXiv:2201.11300}, 2022.

\bibitem{bordenabe2014optimal}
N.~E. Bordenabe, K.~Chatzikokolakis, and C.~Palamidessi, ``Optimal geo-indistinguishable mechanisms for location privacy,'' in \emph{Proceedings of the 2014 ACM SIGSAC conference on computer and communications security}, 2014, pp. 251--262.

\bibitem{qiu2024enhancing}
C.~Qiu, ``Enhancing scalability of metric differential privacy via secret dataset partitioning and benders decomposition,'' in \emph{Proceedings of the Thirty-Third International Joint Conference on Artificial Intelligence}, 2024, pp. 1944--1952.

\bibitem{lin2023geo}
Y.~Lin, ``Geo-indistinguishable masking: enhancing privacy protection in spatial point mapping,'' \emph{Cartography and Geographic Information Science}, vol.~50, no.~6, pp. 608--623, 2023.

\bibitem{Pappachan-EDBT2023}
P.~Pappachan, C.~Qiu, A.~Squicciarini, and V.~S.~H. Manjunath, ``User customizable and robust geo-indistinguishability for location privacy,'' in \emph{EDBT}, 2023.

\bibitem{wang2017location}
L.~Wang, D.~Yang, X.~Han, T.~Wang, D.~Zhang, and X.~Ma, ``Location privacy-preserving task allocation for mobile crowdsensing with differential geo-obfuscation,'' in \emph{Proceedings of the 26th International Conference on World Wide Web}, 2017, pp. 627--636.

\bibitem{wang2019mobile}
L.~Wang, D.~Yang, X.~Han, D.~Zhang, and X.~Ma, ``Mobile crowdsourcing task allocation with differential-and-distortion geo-obfuscation,'' \emph{IEEE Transactions on Dependable and Secure Computing}, vol.~18, no.~2, pp. 967--981, 2019.

\bibitem{qian2021optimal}
Y.~Qian, Y.~Ma, J.~Chen, D.~Wu, D.~Tian, and K.~Hwang, ``Optimal location privacy preserving and service quality guaranteed task allocation in vehicle-based crowdsensing networks,'' \emph{IEEE Transactions on Intelligent Transportation Systems}, vol.~22, no.~7, pp. 4367--4375, 2021.

\bibitem{Qiu-CIKM2020}
C.~Qiu, A.~Squicciarini, Z.~Li, C.~Pang, and L.~Yan, ``Time-efficient geo-obfuscation to protect worker location privacy over road networks in spatial crowdsourcing,'' in \emph{Proceedings of the 29th ACM International Conference on Information \& Knowledge Management}, 2020, pp. 1275--1284.

\bibitem{zhang2022task}
P.~Zhang, X.~Cheng, S.~Su, and N.~Wang, ``Task allocation under geo-indistinguishability via group-based noise addition,'' \emph{IEEE Transactions on Big Data}, vol.~9, no.~3, pp. 860--877, 2022.

\bibitem{qiu2024fine}
C.~Qiu, S.~Yadav, Y.~Ji, A.~C. Squicciarini, R.~Dantu, J.~Zhao, and C.-Z. Xu, ``Fine-grained geo-obfuscation to protect workers' location privacy in time-sensitive spatial crowdsourcing.'' in \emph{EDBT}, 2024, pp. 373--385.

\bibitem{zhang2018shiftroute}
P.~Zhang, C.~Hu, D.~Chen, H.~Li, and Q.~Li, ``Shiftroute: Achieving location privacy for map services on smartphones,'' \emph{IEEE Transactions on Vehicular Technology}, vol.~67, no.~5, pp. 4527--4538, 2018.

\bibitem{yu2023balancing}
D.~Yu, X.~Shi, L.~Chai, W.-A. Zhang, and J.~Chen, ``Balancing localization accuracy and location privacy in mobile cooperative localization,'' \emph{IEEE Transactions on Signal Processing}, vol.~71, pp. 2804--2818, 2023.

\bibitem{min2023geo}
M.~Min, H.~Zhu, S.~Yang, J.~Xu, J.~Tong, S.~Li, and J.~Shu, ``Geo-perturbation for task allocation in 3-d mobile crowdsourcing: An a3c-based approach,'' \emph{IEEE Internet of Things Journal}, vol.~11, no.~2, pp. 1854--1865, 2023.

\bibitem{liu2018epic}
J.~Liu, C.~Zhang, and Y.~Fang, ``Epic: A differential privacy framework to defend smart homes against internet traffic analysis,'' \emph{IEEE Internet of Things Journal}, vol.~5, no.~2, pp. 1206--1217, 2018.

\bibitem{qiu2022trafficadaptor}
C.~Qiu, L.~Yan, A.~Squicciarini, J.~Zhao, C.~Xu, and P.~Pappachan, ``Trafficadaptor: an adaptive obfuscation strategy for vehicle location privacy against traffic flow aware attacks,'' in \emph{Proceedings of the 30th International Conference on Advances in Geographic Information Systems}, 2022, pp. 1--10.

\bibitem{boedihardjo2024metric}
M.~Boedihardjo, T.~Strohmer, and R.~Vershynin, ``Metric geometry of the privacy-utility tradeoff,'' \emph{arXiv preprint arXiv:2405.00329}, 2024.

\bibitem{Shokri-CCS2012}
R.~Shokri, G.~Theodorakopoulos, C.~Troncoso, J.~Hubaux, and J.~L. Boudec, ``Protecting location privacy: Optimal strategy against localization attacks,'' in \emph{Proc. of ACM CCS}, 2012, pp. 617--627.

\bibitem{Reza-PoPET2015}
R.~Shokri, ``Privacy games: Optimal user-centric data obfuscation,'' \emph{PETS}, 2015.

\bibitem{Wang-CIDM2016}
L.~Wang, D.~Zhang, D.~Yang, B.~Y. Lim, and X.~Ma, ``Differential location privacy for sparse mobile crowdsensing,'' in \emph{2016 IEEE 16th International Conference on Data Mining (ICDM)}, 2016, pp. 1257--1262.

\bibitem{Wang-WWW2017}
L.~Wang, D.~Yang, X.~Han, T.~Wang, D.~Zhang, and X.~Ma, ``Location privacy-preserving task allocation for mobile crowdsensing with differential geo-obfuscation,'' in \emph{Proc. of ACM WWW}, 2017, pp. 627--636.

\bibitem{oya2017back}
S.~Oya, C.~Troncoso, and F.~P{\'e}rez-Gonz{\'a}lez, ``Back to the drawing board: Revisiting the design of optimal location privacy-preserving mechanisms,'' in \emph{Proceedings of the 2017 ACM SIGSAC Conference on Computer and Communications Security}, 2017, pp. 1959--1972.

\bibitem{alvim2018metric}
M.~S. Alvim~et al., ``Metric-based local differential privacy for statistical applications,'' \emph{arXiv}, 2018.

\bibitem{Mendes-PETS2020}
R.~Mendes, M.~Cunha, and J.~Vilela, ``Impact of frequency of location reports on the privacy level of geo-indistinguishability,'' \emph{Proceedings on Privacy Enhancing Technologies}, vol. 2020, pp. 379--396, 04 2020.

\bibitem{da2021react}
Y.~Da, R.~Ahuja, L.~Xiong, and C.~Shahabi, ``React: real-time contact tracing and risk monitoring via privacy-enhanced mobile tracking,'' in \emph{2021 IEEE 37th International Conference on Data Engineering (ICDE)}.\hskip 1em plus 0.5em minus 0.4em\relax IEEE, 2021, pp. 2729--2732.

\bibitem{qiu2025time}
C.~Qiu, R.~Liu, P.~Pappachan, A.~Squicciarini, and X.~Xie, ``Time-efficient locally relevant geo-location privacy protection,'' \emph{Proceedings on Privacy Enhancing Technologies}, 2025.

\bibitem{Liu-CCS2025}
R.~Liu and C.~Qiu, ``Panda: Rethinking metric differential privacy optimization at scale with anchor-based approximation,'' in \emph{Proceedings of The 32nd ACM Conference on Computer and Communications Security (CCS)}, 2025.

\bibitem{carvalho2023tem}
R.~S. Carvalho, T.~Vasiloudis, O.~Feyisetan, and K.~Wang, ``Tem: High utility metric differential privacy on text,'' in \emph{Proceedings of the 2023 SIAM International Conference on Data Mining (SDM)}.\hskip 1em plus 0.5em minus 0.4em\relax SIAM, 2023, pp. 883--890.

\bibitem{takagi2019geo}
S.~Takagi, Y.~Cao, Y.~Asano, and M.~Yoshikawa, ``Geo-graph-indistinguishability: Protecting location privacy for lbs over road networks,'' in \emph{Data and Applications Security and Privacy XXXIII: 33rd Annual IFIP WG 11.3 Conference, DBSec 2019, Charleston, SC, USA, July 15--17, 2019, Proceedings 33}.\hskip 1em plus 0.5em minus 0.4em\relax Springer, 2019, pp. 143--163.

\bibitem{zhang2022area}
P.~Zhang~et al., ``Area coverage-based worker recruitment under geo-indistinguishability,'' \emph{Computer Networks}, 2022.

\bibitem{meisenbacher20241}
S.~Meisenbacher~et al., ``Efficient and utility-preserving text obfuscation leveraging word-level metric differential privacy,'' \emph{arXiv}, 2024.

\bibitem{yan2024coder}
H.~Yan, X.~Li, W.~Zhang, Q.~Chen, B.~Wang, H.~Li, and X.~Lin, ``Coder: Protecting privacy in image retrieval with differential privacy,'' \emph{IEEE Transactions on Dependable and Secure Computing}, 2024.

\bibitem{zheng2010geolife}
Y.~Zheng, L.~Zhang, X.~Xie, and W.-Y. Ma, ``Mining interesting locations and travel sequences from gps trajectories,'' in \emph{Proceedings of the 19th international conference on World wide web}.\hskip 1em plus 0.5em minus 0.4em\relax ACM, 2010, pp. 791--800.

\bibitem{mendes2020impact}
R.~Mendes, M.~Cunha, and J.~P. Vilela, ``Impact of frequency of location reports on the privacy level of geo-indistinguishability,'' \emph{Proceedings on Privacy Enhancing Technologies}, 2020.

\bibitem{yuan2010t}
J.~Yuan, Y.~Zheng, C.~Zhang, W.~Xie, X.~Xie, G.~Sun, and Y.~Huang, ``T-drive: driving directions based on taxi trajectories,'' in \emph{Proceedings of the 18th SIGSPATIAL International conference on advances in geographic information systems}, 2010, pp. 99--108.

\bibitem{cho2011gowalla}
E.~Cho, S.~A. Myers, and J.~Leskovec, ``Friendship and mobility: User movement in location-based social networks,'' in \emph{Proceedings of the 17th ACM SIGKDD international conference on Knowledge discovery and data mining}.\hskip 1em plus 0.5em minus 0.4em\relax ACM, 2011, pp. 1082--1090.

\bibitem{oya2017geo}
S.~Oya, C.~Troncoso, and F.~P{\'e}rez-Gonz{\'a}lez, ``Is geo-indistinguishability what you are looking for?'' in \emph{Proceedings of the 2017 on Workshop on Privacy in the Electronic Society}, 2017, pp. 137--140.

\bibitem{yang2013foursquare}
D.~Yang, D.~Zhang, V.~W. Liu, and H.~Zha, ``A categorial and geographical-aware recommender system,'' in \emph{Proceedings of the 22nd International Conference on World Wide Web}.\hskip 1em plus 0.5em minus 0.4em\relax ACM, 2013, pp. 1447--1458.

\bibitem{wang2018personalized}
Z.~Wang, J.~Hu, R.~Lv, J.~Wei, Q.~Wang, D.~Yang, and H.~Qi, ``Personalized privacy-preserving task allocation for mobile crowdsensing,'' \emph{IEEE Transactions on Mobile Computing}, vol.~18, no.~6, pp. 1330--1341, 2018.

\bibitem{piorkowski2009cabspotting}
M.~Piorkowski, N.~Sarafijanovic-Djukic, and M.~Grossglauser, ``Hail cabs in the air: A large-scale study of gps-enabled taxis,'' in \emph{Proceedings of the 9th ACM SIGCOMM conference on Internet measurement conference}, 2009, pp. 386--398.

\bibitem{moreira2013portocabs}
L.~Moreira-Matias, J.~Gama, M.~Ferreira, J.~Mendes-Moreira, and L.~Damas, ``Predicting taxi-passenger demand using streaming data,'' in \emph{IEEE Transactions on Intelligent Transportation Systems}, vol.~14, no.~3.\hskip 1em plus 0.5em minus 0.4em\relax IEEE, 2013, pp. 1393--1402.

\bibitem{harper2015movielens}
F.~M. Harper and J.~A. Konstan, ``The movielens datasets: History and context,'' \emph{Acm transactions on interactive intelligent systems (tiis)}, vol.~5, no.~4, pp. 1--19, 2015.

\bibitem{yelp_dataset_challenge}
{Yelp Inc.}, ``{Yelp Dataset},'' \url{https://www.yelp.com/dataset}, 2024, accessed: [Your Access Date].

\bibitem{cohen2017emnist}
G.~Cohen, S.~Afshar, J.~Tapson, and A.~Van~Schaik, ``Emnist: Extending mnist to handwritten letters,'' in \emph{2017 international joint conference on neural networks (IJCNN)}.\hskip 1em plus 0.5em minus 0.4em\relax IEEE, 2017, pp. 2921--2926.

\bibitem{maas2011imdb}
A.~L. Maas, R.~E. Daly, P.~T. Pham, D.~Huang, A.~Y. Ng, and C.~Potts, ``Learning word vectors for sentiment analysis,'' in \emph{Proceedings of the 49th annual meeting of the Association for Computational Linguistics: Human Language Technologies}, 2011, pp. 142--150.

\bibitem{potthast2011pan}
M.~Potthast, B.~Stein, A.~Barr{\'o}n-Cede{\~n}o, and P.~Rosso, ``Overview of the 3rd international competition on plagiarism detection,'' in \emph{Notebook for PAN at CLEF 2011}, 2011.

\bibitem{stamatatos2012pan}
E.~Stamatatos, M.~Kestemont, B.~Stein, and M.~Potthast, ``Overview of the pan 2012 traditional authorship attribution task,'' in \emph{CLEF (Online Working Notes/Labs/Workshop)}, 2012.

\bibitem{pang2005seeing}
B.~Pang and L.~Lee, ``Seeing stars: Exploiting class relationships for sentiment categorization with respect to rating scales,'' in \emph{Proceedings of the 43rd annual meeting of the Association for Computational Linguistics (ACL'05)}, 2005, pp. 115--124.

\bibitem{hu2004mining}
M.~Hu and B.~Liu, ``Mining and summarizing customer reviews,'' in \emph{Proceedings of the tenth ACM SIGKDD international conference on Knowledge discovery and data mining}, 2004, pp. 168--177.

\bibitem{simplemaps_uscities}
{SimpleMaps}, ``{U.S. Cities Database},'' \url{https://simplemaps.com/data/us-cities}, 2023, a commonly used commercial/free database for U.S. city locations and populations.

\bibitem{uscensusbureau_acs}
{U.S. Census Bureau}, ``{American Community Survey (ACS)},'' \url{https://www.census.gov/programs-surveys/acs}, 2023, ongoing survey, year of data used should be specified.

\bibitem{twitter_api}
{Twitter, Inc. (now X Corp.)}, ``{Twitter API},'' \url{https://developer.twitter.com/}, 2023, data is collected via the official API; specific datasets are usually not public.

\bibitem{yang2016revisiting}
Z.~Yang, W.~W. Cohen, and R.~Salakhutdinov, ``Revisiting semi-supervised learning with graph embeddings,'' in \emph{International conference on machine learning}.\hskip 1em plus 0.5em minus 0.4em\relax PMLR, 2016, pp. 40--48, this paper popularized the use of citation network datasets like Cora, CiteSeer, and DBLP for graph-based learning tasks.

\bibitem{openstreetmap_foundation}
{OpenStreetMap contributors}, ``{OpenStreetMap},'' \url{https://www.openstreetmap.org}, 2024, data licensed under the Open Data Commons Open Database License (ODbL).

\bibitem{balk2004grump}
D.~L. Balk, F.~Pozzi, G.~Yetman, U.~Deichmann, and A.~Nelson, ``{GRUMPv1: Global Rural-Urban Mapping Project, Version 1},'' Columbia University. Palisades, NY, 2004, \url{https://sedac.ciesin.columbia.edu/data/set/grump-v1-population-density}.

\bibitem{google_places_api}
{Google}, ``{Google Places API},'' \url{https://developers.google.com/maps/documentation/places}, 2024, part of the Google Maps Platform.

\bibitem{elsalamouny2014generalized}
E.~ElSalamouny, K.~Chatzikokolakis, and C.~Palamidessi, ``Generalized differential privacy: Regions of priors that admit robust optimal mechanisms,'' \emph{Horizons of the Mind. A Tribute to Prakash Panangaden: Essays Dedicated to Prakash Panangaden on the Occasion of His 60th Birthday}, pp. 292--318, 2014.

\bibitem{elsalamouny2016differential}
E.~ElSalamouny~et al., ``Differential privacy models for location-based services,'' \emph{Transactions on Data Privacy}, 2016.

\bibitem{odoh2024group}
K.~Odoh, ``Group-wise k-anonymity meets ($\varepsilon$, $\delta$) differentially privacy scheme,'' in \emph{Companion Proceedings of the ACM Web Conference 2024}, 2024, pp. 802--805.

\end{thebibliography}
\end{document}